\documentclass[%
    reprint,
    showpacs,preprintnumbers,
    showkeys,
    amsmath,amssymb,
    aps,
    pra,
    floatfix,
    longbibliography,
    notitlepage,
]{revtex4-2}

\usepackage[english]{babel}
\usepackage[utf8]{inputenc}
\usepackage{graphicx}
\usepackage{bm}
\usepackage[T1]{fontenc}
\usepackage{hyperref}
\usepackage{xcolor}
\usepackage{epsfig}
\usepackage{amsmath}
\usepackage{amsfonts}
\usepackage{amssymb}
\usepackage{mathtools}
\usepackage{bbold}
\usepackage{amsthm}
\usepackage[caption=false]{subfig}
\usepackage{gensymb}
\usepackage{enumerate}
\usepackage{tikz}
\usepackage{dsfont}
\usepackage[all]{nowidow}
\usepackage{pifont}
\usepackage{algorithm}
\usepackage{algpseudocode}
\usepackage[capitalise]{cleveref}
\usepackage[export]{adjustbox}
\usepackage{xspace}
\usepackage{nicefrac}

\definecolor{refColor}{HTML}{0376E9}
\definecolor{figColor}{HTML}{0376E9}
\definecolor{urlColor}{HTML}{0376E9}

\hypersetup{%
    unicode=false,                 
    pdftoolbar=true,               
    pdfmenubar=true,               
    pdffitwindow=false,            
    pdfstartview={FitH},           
    pdfpagemode=false,             
    pdftitle={},                   
    pdfauthor={Nadine Lenke},      
    pdfsubject={Quantum physics},  
    pdfcreator={Guido Burkard},     
    pdfproducer={Nadine Lenke},    
    pdfkeywords={},                
    pdfnewwindow=true,             
    colorlinks=true,               
    linkcolor=figColor,            
    citecolor=refColor,            
    filecolor=magenta,             
    urlcolor=urlColor              
}

\begin{document}

\title{Reconstruction of the noise correlation spectral density \\ from the cavity emission in a two-qubit system}

\author{Nadine Lenke}
\author{Guido Burkard}
\affiliation{%
    Department of Physics, University of Konstanz, D-78457 Konstanz, Germany
}



\begin{abstract}
   A significant challenge in the field of large-scale fault-tolerant quantum computation is the influence of noise. In addition to the influence of noise on individual qubits, the smaller additional effect of noise correlations is also of high significance because correlated errors pose a challenge for quantum error correction. We describe the dynamics of two cavity-coupled qubits that are subject to correlated noise, assuming that the qubits are affected by longitudinal noise and not coupled directly. We find that the cavity emission enables the characterization of the noise correlations and describe the cases of white noise, quasi-static noise, and Ornstein-Uhlenbeck noise. For a known frequency spectrum, the reconstruction of the noise correlation spectral density from the cavity emission is possible by averaging over many different noise realizations. We demonstrate that, in the case of white noise, the noise correlation effects scale with the fifth order of the cavity-qubit coupling constant and are thus strongly suppressed compared with the case of quasi-static noise, where they scale with the third order.  Furthermore, we present a method for extracting the noise correlation spectral density from the cavity emission in the case where the underlying noise type remains unidentified. This can be achieved by applying the convolution theorem. 
\end{abstract}

\maketitle

\section{Introduction}
Current research is increasingly focused on realizing large-scale quantum processors. Such systems are required to implement quantum algorithms while maintaining fault tolerance. Several physical platforms have been proposed for  qubit realization, including superconducting circuits \cite{decosupersuperqubit,supersupersinglequbitgate,supersupertwoqubitgate,supersuperreadout,supersuperquantumerrorcorrection} and trapped-ion qubits \cite{superionqubitslifetime,superionqubitssinglequibitgates,superionqubittwoqubitgate}. Among these, semiconductor spin qubits are considered particularly promising due to their long coherence times, compatibility with established semiconductor fabrication techniques, and the potential for high-density integration \cite{veldhorstsuperqubit,muhonensuperqubit,taruchasuperqubit,readoutfidelitysilicon,quantumerror}. Detailed reviews on various qubit platforms can be found in  \cite{guidostutorial,guidostutorialvon2020}. However, scaling spin-qubit architectures to large numbers of qubits remains a major challenge, as larger systems become increasingly susceptible to decoherence and operational errors caused by various noise mechanisms \cite{noisedephasing, thermalnoise}. In particular, correlated noise processes represent a detrimental challenge, because they significantly reduce single- and two-qubit gate fidelities and limit the performance of quantum error correction  \cite{noicorrtheoryloss}.

To realize scalable quantum computation, fast and high-fidelity single- and two-qubit operations are essential. In small systems, two qubit operations are typically realized from next-neighbour exchange coupling. Spin qubits exhibit long coherence times due to their weak coupling to the environment, but this property makes long-range qubit interactions, which are required for scaling up, more challenging. A possibility for realizing such interactions is provided by cavity quantum electrodynamics (cQED), where qubits are embedded in a microwave cavity that enables entanglement through the exchange of photons over large distances. This interaction additionally allows measurements of the spin dynamics using high-fidelity dispersive readout \cite{pettaletter2012, circuitqed21}. Superconducting microwave resonators lend themselves to very sensitive detectors for quantum sensing and single-photon detection \cite{,Beaulieu2025,Oppliger2026}. The coupling between a spin qubit and the cavity is mediated by the electric dipole interaction between cavity photons and the charge degree of freedom of the qubit.  Realizations of the cavity-mediated coupling of two spin qubits were proposed in \cite{benito2019cavity} and experimentally demonstrated \cite{dijkema2025cavity,  2020naturepetta}. However, the performance of such cavity-mediated architectures is ultimately limited by noise, which becomes particularly relevant in the context of scalable spin-qubit implementations.

Currently, we are in the era of NISQ (noisy intermediate-scale quantum) systems, where noise limits the performance and control of quantum processors. Therefore, understanding the noise becomes essential for scaling of quantum circuits. The influence of noise introduces readout errors \cite{readouterror} and has a significant impact on dephasing, with this phenomenon mainly influenced by low-frequency noise, such as $1/f$-noise \cite{noisedephasing}. Additionally, it limits the performance of two-qubit gates \cite{twoqubitgatesinfluenceofnoise, decosupersuperqubit}. The most common noise sources affecting semiconductor spin qubits are charge noise \cite{guidostutorial, thermalnoise}, spin noise \cite{semichargespin}, and magnetic-field drifts \cite{losspaper2026}.  A wide variety of theoretical descriptions of the influence of  1/f noise \cite{thermalnoise}, Ornstein-Uhlenbeck noise \cite{ounoisequelle,ounoisequelle2}, measurement-induced shot noise \cite{noisesimulationsantos}, and non-Gaussian noise \cite{nongaussiannoisequelle} on a single qubit or multiple qubits have already been studied. The signatures of Gaussian noise on a qubit embedded in a cavity in the cavity transmission was theoretically investigated in \cite{pm,pm2}. In the aforementioned studies, the presence of noise in the energy separation of the qubit was compared to the influence of noise on the cavity-qubit coupling $g$.  However, in larger qubit arrays, noise effects become more severe due to additional noise correlation effects between qubits. These correlations strongly limit scalability by increasing crosstalk and correlated errors \cite{noicorrtheoryloss,correlatederrorsquelle1}. Since quantum error correction (QEC) can be used to achieve fault tolerance, its performance is particularly sensitive to such correlated noise.
 
In particular, quantum error correction (QEC) \cite{qecbeginners,qecguide} performs significantly worse in the presence of correlated noise, since it induces correlated errors.  Most QEC protocols assume weak uncorrelated noise, which leads to local single-qubit errors that can be corrected step by step \cite{qecforarbitrarynoise,qeceffectsalgorithms,qeccorrelations}. In contrast, correlated noise severely degrades QEC performance: error rates become harder to estimate, and only a limited subset of errors can be corrected. Accurate QEC therefore requires additional correction steps and substantial qubit overhead  \cite{losspaper2026,QECreweighting,qecbarbara,qeccosmic}. 
Correlated noise also limits QEC scalability, as the suppression of logical error rates with increasing code distance is much weaker than for uncorrelated noise, requiring larger error-correcting codes \cite{qeccorrelations,qechamiltonian, losspaper2026}. Furthermore, correlated errors may occur simultaneously or within short time frames, while QEC codes typically correct only one error per cycle \cite{losspaper2026,qechamiltonian}. Since QEC relies on redundancy by comparing measurements across many qubits, simultaneous errors reduce the effective number of logical qubits \cite{losspaper2026,QECreweighting}.
The impact of correlated noise depends on its source. Magnetic field drifts \cite{losspaper2026} and correlated phase errors \cite{qecphaseerror} are particularly challenging.

So far, most research has focused on describing noise effects on individual qubits, while the additional effects of noise correlations remain challenging both theoretically and experimentally. In this regard, experiments were performed on qubits formed in quantum dots in Si/SiGe to determine noise correlations in the presence of charge noise or nuclear spin noise \cite{noicorrinsisige,noicorrpairofqubits,dijkema,chargenoisemeasure}. Correlated errors can result from cosmic rays, as observed in an array consisting of superconducting qubits \cite{qeccosmic}. Theoretical descriptions of noise correlations in a system of two qubits are presented in \cite{noicorrtheoryloss}, where the influence of spatio-temporally correlated 1/f noise is examined in different regimes. In \cite{losspaper2026} correlated noise originating from magnetic fields and charge noise is examined in a system containing multiple qubits. Characterization of noise affecting two qubits can be achieved using the method proposed in \cite{rojasnoicorrloss25}, where single-qubit Ramsey sequences are applied to the two qubits, such that the noise spectral densities can be recovered over a broad frequency range.

In this regard, a general theoretical description of temporal noise correlations poses a challenge. So far, most research has been done for particular types of noise or for cases where a special technique is required. In this paper, our aim is to provide a broadly applicable theoretical description of noise correlations using the methods of cQED. We examine the influence of correlated noise following a multivariate Gaussian distribution on the energy separations of two qubits. The influence of the noise is characterized indirectly through the emission of a cavity to which the qubits are coupled. The qubits are placed in a sufficiently large distance from each other such that the direct coupling between them can be neglected and only the indirect effects of the coupling to the same cavity are relevant. For white noise, we find that noise correlations do not affect the cavity emission up to third order in the qubit–cavity coupling strength. In the case of quasi-static noise, we find a contribution of noise correlations to the cavity emission that depends approximately linearly on the noise-correlation spectral density amplitude. Ornstein-Uhlenbeck noise represents an intermediate case between white noise and quasi-static noise, since in the limit of small noise correlation decay rates, the change in the cavity output field originating from the noise correlations is linear, while in the limit of large noise correlation decay rates, the effect of the noise correlations is washed out. For general noise, without assuming a specific noise model, the noise-correlation spectral density can still be extracted from the cavity emission using the convolution theorem.

This paper is structured as follows: In Sec.~\ref{sec:model_system}, we introduce the underlying system and describe the dynamics of the qubit state and the cavity. The noise models are presented in Sec.~\ref{sec:model_noise}. In Sec.~\ref{sec:cavityemission}, we analyze the cavity emission by solving the quantum Langevin equations and derive expressions for the output field both in the noiseless case and in the presence of noise, which is averaged over many measurements. The effects of noise correlations on the cavity emission are then investigated for white noise (Sec.~\ref{sec:white}), quasi-static noise (Sec.~\ref{sec:quasi}), and Ornstein–Uhlenbeck noise (Sec.~\ref{sec:ou}). In Sec.~\ref{sec:general}, we introduce a method for extracting the noise-correlation spectral density in the general case, where the noise type and frequency dependence are unknown. Finally, the main results are summarized in Sec.~\ref{sec:conclusions}.
\section{Model: System}
\label{sec:model_system}
\begin{figure}
    \centering
    \includegraphics[width=0.95\linewidth]{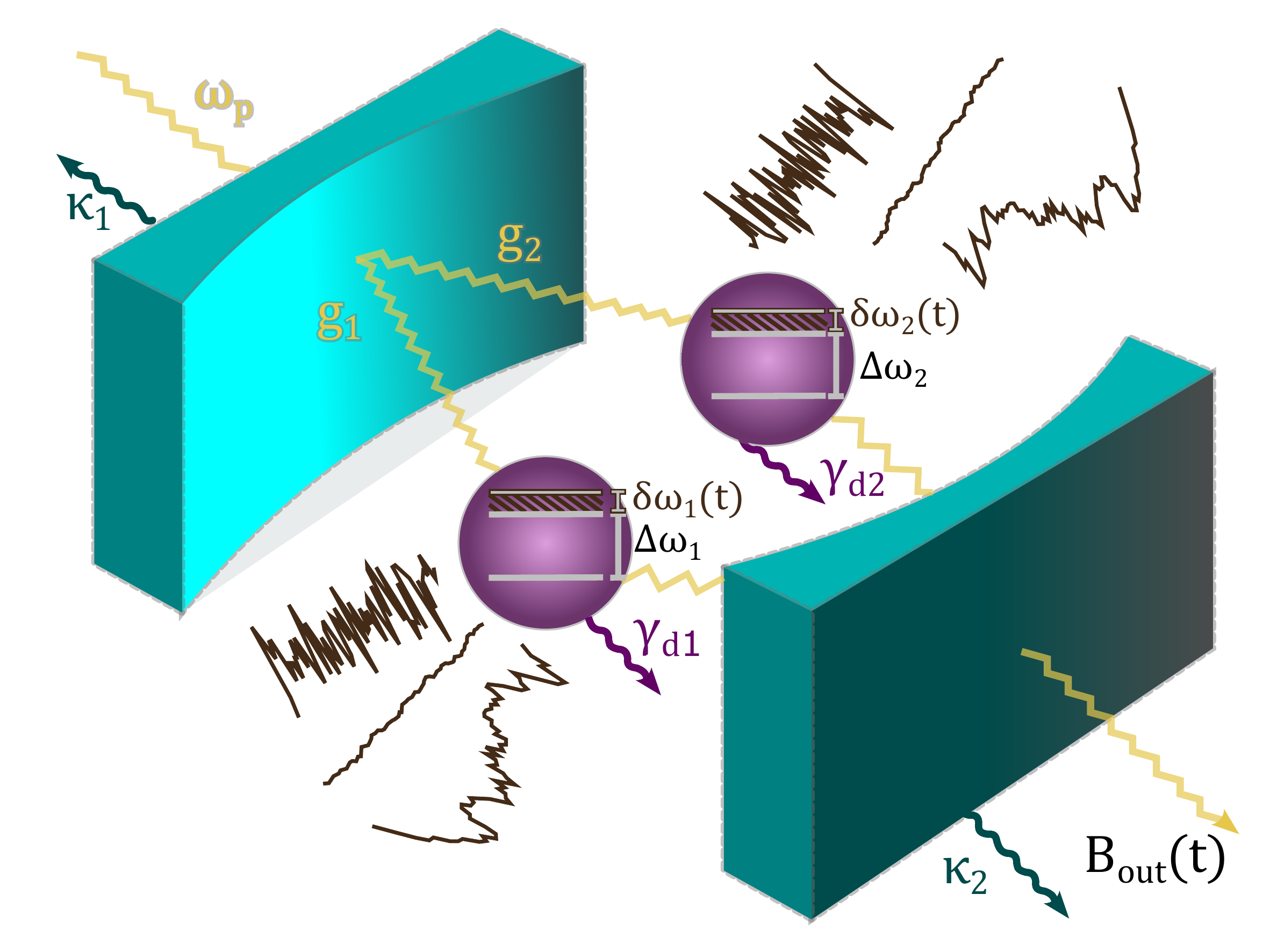}
    \caption{The system under consideration consists of a cavity (turquoise), here schematically represented by two mirrors, and two qubits ($k=1,2$; purple), each of which is coupled to the cavity mode of frequency $\omega_c$ with transversal coupling strength $g_k$. The qubits are arranged such that there is no direct correlation or coupling between them. We aim to detect indirect correlations between the qubits caused by correlated noise (brown) through the emission from the cavity. The qubits undergo additional dephasing processes with rates $\gamma_{\phi k}$ and relaxation processes with rates $\gamma_k$. Both are captured in the total decoherence rate $\gamma_{dk}=\gamma_k+2\gamma_{\phi k}$. The cavity loss rates at the two ports are denoted $\kappa_1$ and $\kappa_2$. The rotating frame corotates with the probe frequency of the laser, denoted by $\omega_p$. An output field $B_{\rm out}(t)$ can be detected on the second cavity port, even in the absence of an input field. This arises from the decay process of the qubits into their ground states. The energy difference between the qubit ground and excited states is given by $\Delta\omega_k$, or alternatively, by the detunings $\Delta_k = \Delta\omega_k-\omega_p$. Both qubits are affected by time-varying noise sources, $\delta \omega_{k}(t)$. }
    \label{setup_figure}
\end{figure}

The system we are describing consists of two qubits labeled $k=1,2$ that are coupled to the same mode of an electromagnetic cavity.  In the case of semiconductor spin qubits or superconducting qubits, the cavity will typically be a superconducting co-planar waveguide cavity. We assume that both qubits are subject to longitudinal noise that affects their energy splitting $\Delta\omega_{k}$. The emission of a photon from the qubits contributes to the cavity emission, denoted by $B_{\text{out}}(t)$. The cavity is subject to losses, where the loss rates at the cavity ports are described by the rates $\kappa_1$ and $\kappa_2$. The configuration of the system is illustrated in Fig.~\ref{setup_figure}.

The coupled qubit-cavity system can be described by the Hamiltonian, 
\begin{subequations}
\allowdisplaybreaks
    \label{hamiltonian_komplett}
    \begin{align}
         \mathcal{H}(t)&=\hat{H}_\text{sys}(t)+\hat{H}_{\text{bath}}+\hat{H}_{\text{int}}, \tag{1}\\
        \hat{H}_{\text{sys}}(t)&=\Delta_c \hat{a}^\dagger\hat{a}+\sum_{k=1,2}\bigg[\frac{\Delta_k+\delta\omega_{k}(t)}{2}\hat{\sigma}_{zk}\notag  \\
        &\quad +g_k\left( \hat{\sigma}_{+k}\hat{a}+\hat\sigma_{-k}\hat{a}^\dagger\right)\bigg],\label{1a}\\
        \hat{H}_{\text{bath}}&=\sum_{j=1,2}\int_{-\infty}^\infty d\omega  \omega \hat{b}_j^\dagger(\omega)\hat{b}_j(\omega), \label{1b}\\
         \hat{H}_{\text{int}}&=\sum_{j=1,2}\int_{-\infty}^\infty d \omega  i\tilde{\kappa}_j(\omega)\left( \hat{b}_j(\omega)\hat{a}^\dagger-\hat{b}_j^\dagger(\omega)\hat{a}\right), \label{1c}
    \end{align}    
\end{subequations}
within the rotating frame and after the rotating wave approximation was performed, for details, see Appendix~\ref{sec:apphamil}. The system part of the Hamiltonian (Eq.~\eqref{1a}) describes the dynamics of the qubit, including the influence of noise and the photon exchange between the qubits and the cavity. The bath part corresponds to the photons of the environment (Eq.~\eqref{1b}) and the interaction part of the Hamiltonian, labeled with ``int'' expresses the interaction between the environment and the cavity (Eq.~\eqref{1c}). The cavity-probe detuning $\Delta_c=\omega_c-\omega_p$ is introduced as the difference between the cavity mode frequency $\omega_c$ and the probe frequency $\omega_p$. The creation and annihilation operators for photons in the cavity are denoted by $\hat{a}^\dagger$ and $\hat{a}$, respectively. The index $k\in\{1,2\}$ refers to the two qubits, while the index $j \in \{1,2\} $ labels the two ports of the cavity. The qubit-probe detunings for the respective qubits $k$ are introduced by $\Delta_k=\Delta \omega_{k}-\omega_p$, where $\Delta \omega_{k}$ refers to the energy level spacings in the laboratory frame.  They are affected by longitudinal time-dependent noise $\delta \omega_{k}(t)$, which will be described in detail in Sec.~\ref{sec:model_noise}. The terms including Pauli matrices $\hat{\sigma}_{zk}$ and $\hat{\sigma}_{\pm k}=(\hat{\sigma}_{xk}\pm i \hat{\sigma}_{yk})/2$ describe the state of the qubits. The transverse coupling between the qubit $k$ and the cavity is described with the coupling constant $g_k$. The frequency dependent operators $\hat{b}_j^\dagger(\omega)$ and $\hat{b}_j(\omega)$ describe the creation and annihilation of photons outside the cavity ports. Finally, $\tilde{\kappa}_j(\omega)$ is the coupling strength between the cavity ports and the thermal bath, i.e., the surroundings of the cavity. Note that we use units in which $\hbar=1$ throughout this work. For details of the derivation of the Hamiltonian, see App.~\ref{sec:apphamil}.

From this Hamiltonian, we can derive the Lindblad master equation, which describes the dynamics of the system with the additional influence of dissipative processes \cite{inoutgardiner}. Emission processes are described with the dissipators $\hat{\sigma}_{-k}$, where the influences of spontaneous and stimulated emission with their respective loss rates $\gamma_k$ and $n\gamma_k$ are taken into account. Absorption processes with rates $n\gamma_k$ and dissipators $\hat{\sigma}_{+k}$ are also considered, as well as qubit dephasing processes, which can be modelled with dissipators $\hat{\sigma}_{zk}/\sqrt{2}$ and the corresponding rates of $\gamma_{\phi k}$. Here, $\gamma_k$ is the qubit relaxation rate, $\gamma_{\phi k}$ denotes the qubit dephasing rate, and $n_k$ is the average occupation number of the thermal bath $n_k=\left[\text{exp}(\Delta\omega_{k}/T)-1\right]^{-1}$ evaluated at the qubit energy splitting where resonant energy exchange is possible \cite{pm}.  The Lindblad master equation, which describes the qubit state dynamics, can be stated as 
\begin{equation}
    \label{quantum_master_equation}
    \begin{aligned}
        \frac{d\hat{\rho}}{dt}&=i\left[ \hat{\rho},\mathcal{H}(t)\right]+\sum_{k=1,2}(1+n_k)\gamma_k \mathcal{D}\left[ \hat{\sigma}_{-k}\right]\hat{\rho}\\
        &\quad +n_k\gamma_k\mathcal{D}\left[\hat{\sigma}_{+k}\right] \hat{\rho}+\frac{\gamma_{\phi k}}{2}\mathcal{D}\left[\hat{\sigma}_{zk}\right]\hat{\rho},
    \end{aligned}
\end{equation}
where $\rho$ denotes the density operator of the two qubits and the dissipators are defined as $\mathcal{D}\big[ \hat{L}\big]\hat{\rho}=\hat{L}\hat{\rho}\hat{L}^\dagger-\frac{1}{2}\hat{L}^{\dagger}\hat{L}\hat{\rho}-\frac{1}{2}\hat{\rho}\hat{L}^{\dagger}\hat{L}$.

This master equation enables the formulation of the equations of motion for the expectation values of operators $\hat{\mathcal{O}}$ in the Schrödinger picture, according to the relations $\langle\hat{\mathcal{O}}\rangle=\text{Tr}\big[\hat{\mathcal{O}}\hat{\rho}\big]$ and $d\langle\hat{\mathcal{O}}\rangle/dt=\text{Tr}\left(\hat{\mathcal{O}}\big(d\hat{\rho}/dt\big)\right)$. These will then describe the dynamics of the system, 
\begin{equation}
    \label{quantum_master_equation}
    \begin{aligned}
    \frac{d\langle\hat{\mathcal{O}}\rangle}{dt}= & i\big\langle\big[\hat{\mathcal{H}}(t),\hat{\mathcal{O}} \big]\big\rangle
        +\sum_{k=1,2}(1+n_k)\gamma_k \langle\bar{\mathcal{D}}\left[ \hat{\sigma}_{-k}\right]\hat{\mathcal{O}}\rangle\\
        &\quad +n_k\gamma_k\langle\bar{\mathcal{D}}\left[\hat{\sigma}_{+k}\right] \hat{\mathcal{O}}\rangle+\frac{\gamma_{\phi k}}{2}\langle\bar{\mathcal{D}}\left[\hat{\sigma}_{zk}\right]\hat{\mathcal{O}}\rangle,
    \end{aligned}
\end{equation}
where $\bar{\mathcal{D}}\big[ \hat{L}\big]\hat{\mathcal{O}}=\hat{L}^\dagger\hat{\mathcal{O}}\hat{L}-\frac{1}{2}\hat{L}^{\dagger}\hat{L}\hat{\mathcal{O}}-\frac{1}{2}\hat{\mathcal{O}}\hat{L}^{\dagger}\hat{L}$.

\section{Model: Noise}
\label{sec:model_noise}
We consider temporally correlated, longitudinal, and classical noise on the energy separations of the two qubits $\Delta\omega_k\rightarrow\Delta\omega_k+\delta\omega_{k}(t)$ for $k=1,2$. The noise influence originates from time-dependent fluctuations $X_k(t)$ in the environment of the two qubits. We assume that the fluctuations have zero mean, $\langle X_k(t)\rangle=0$. The relation between these fluctuations and the noise on the qubits is given by 
\begin{equation}
    \label{linearized_noise_lambda}
    \delta \omega_k(t)=\lambda_k  X_k(t),
\end{equation}
where $\lambda_k$ is the sensitivity to noise for the respective qubit, $\lambda_k=\partial \omega_k/\partial X_k\big|_{ X_k=0}$ \cite{ithier2005}. A technique for measuring this noise sensitivity to the environment is demonstrated in \cite{qubitasprobeqs}. In the case where $\lambda_k=0$, the noise-free situation is recovered. Here, we are interested in the case where the noise that affects the two qubits is correlated. We want to determine whether such noise correlations can be detected in the  emission from a cavity to which the two noise-prone qubits are coupled. The extraction of the frequency spectrum of the noise correlation spectral density $S_{12} (\omega)$, which will be introduced below, allows the differentiation of distinct noise sources from this emission. These noise sources can be described, e.g., as white noise (Sec.~\ref{sec:white}), quasi-static noise (Sec.~\ref{sec:quasi}), or  Ornstein-Uhlenbeck noise (Sec.~\ref{sec:ou}). We model the noise affecting the two qubits as a pair of random time-dependent functions, $X_1(t)$ and $X_2(t)$, with a multivariate Gaussian distribution with zero mean.
The variances of the noise on the first and second qubits are given by $\langle\langle X_1^2\rangle\rangle$ and $\langle\langle X_2^2\rangle\rangle$ respectively, while $\langle\langle X_1 X_2\rangle\rangle$ gives the covariance between both random variables. Since the random variables are assumed to be real-valued, we have  $\langle\langle X_1 X_2\rangle\rangle = \langle\langle X_2 X_1\rangle\rangle$. A valid covariance matrix must be positive definite, which is equivalent to the requirement $\text{det}(\langle\langle X_k X_{k'}\rangle\rangle)>0$ (since $\langle\langle X_1^2\rangle\rangle \ge 0$ per definition). The joint probabilities for the random variables $X_k$ as well as the sum over both, $X_1+X_2$, are also Gaussian distributed with zero mean, as shown in   
\cite{wkeitstheoklenkeeng}.

Now, we are interested in the stochastic phases $\chi_k(t)$ originating from the noise on the qubits, which are defined with the time integral over the noise on the qubits energy separations,
\begin{equation}
    \label{noise_with_lambda}
    \chi_k(t)=\int_{0}^t dt' X_{k}(t') .
\end{equation}
The stochastic phases can be regarded as the sums of many random variables $X_k(t)$ at different times. When the auto-correlators $\langle\langle  X_k(0)  X_k(\tau)\rangle\rangle$ decay on timescales smaller than the integration time $t$, the stochastic phases become sums over uncorrelated Gaussian random variables with zero mean. In this case, the central limit theorem can be applied, stating that the stochastic phases $\chi_k(t)$ will also be Gaussian distributed with zero mean \cite{pm,Bergli,chirolli}.

The second moment of the random variables can be related to the auto-correlator $W_k(|t_1-t_2|)=W_{kk}(|t_1-t_2|)$ where $k=1,2$, or to the cross-correlator $W_{12}(|t_1-t_2|)$ for different time points $t_1,t_2$ \cite{Bergli, rauschespektraldichtealt, wieneralt},
\begin{equation}
    \label{auto_and_cross_correlator_w_def}
    \begin{aligned}
    \langle \langle \chi_k(t')&\chi_{k'}(t)\rangle\rangle
    =\int_0^{t'}dt_1\int_0^{t}dt_2 W_{kk'}(|t_1-t_2|)\\
    &=\int_0^{t'} dt_1\int_0^{t}dt_2\langle\langle  X_{k}(t_1)X_{k'}(t_2)\rangle\rangle .
    \end{aligned}
\end{equation}
Note that $\langle\langle\dots \rangle\rangle$  is used to indicate the average of a quantity in a number of measurements. In addition, the correlators   $W_{kk'}(t)$ are symmetric functions in time, $W_{kk'}(-t)=W_{kk'}(t)$, by definition. More accessible physical quantities are the noise spectral densities $S_k(\omega)=S_{kk}(\omega)$ and the noise correlation spectral density $S_{12}(\omega)$,
\begin{equation}
    \label{auto_and_corr_s_w_def}
    \begin{aligned}
        S_{kk'}(\omega)&=\frac{1}{\pi}\int_0^\infty dt W_{kk'}(t)\text{cos}(\omega t),\\
        W_{kk'}(t)&=\int_{-\infty}^\infty d\omega S_{kk'}(\omega)e^{-i\omega t},
    \end{aligned}
\end{equation}
which are related through a Fourier transform to the correlators.
Note, that the cross-correlator and the noise-correlation spectral density are symmetric under the exchange of indices, $W_{12}(t)=W_{21}(t)$ and $S_{12}(\omega)=S_{21}(\omega)$, which leads to a symmetric noise correlator, $\langle\langle\chi_1(t')\chi_2(t)\rangle\rangle=\langle\langle\chi_2(t)\chi_1(t')\rangle\rangle$. We note that the correlators and the noise spectral densities are real-valued, as is the case for classical noise.

The relation between the second moments of the stochastic phases for different time points and the noise spectral density can be obtained by combining Eqs.~\eqref{auto_and_cross_correlator_w_def} and \eqref{auto_and_corr_s_w_def},
\begin{equation}
    \label{chi1_k_chikp_tp_von_s}
    \begin{aligned}
    \langle\langle \chi_k(t')\chi_{k'}(t)\rangle\rangle&=\int_{-\infty}^\infty 
     \!\!\!\! d\omega \frac{S_{kk'}(\omega)}{\omega^2}\big(e^{i\omega t}-1\big)
     \!\big(e^{-i\omega t'}-1\big).
    \end{aligned}
\end{equation}
This quantity describes the auto- or cross-correlations that contribute to the cavity emission. In the auto-correlation case ($k=k'$) this relation reduces to \cite{Bergli,ithier2005},
\begin{equation}
    \label{chi_1_quadrat_t_von_s}
    \langle\langle\chi_k^2(t)\rangle\rangle=4\int_{-\infty}^\infty d\omega \frac{\text{sin}^2(\omega t/2)}{\omega^2}S_k(\omega),
\end{equation}
for $t=t'$. The emission from the cavity can be derived by solving the integrals in Eqs.~\eqref{chi1_k_chikp_tp_von_s} and \eqref{chi_1_quadrat_t_von_s} which is possible if the frequency dependence of the noise spectral density is known when a particular type of noise is considered. Aiming at a comparison of different noise examples, we will derive the noise influence on the cavity emission in the presence of white noise, quasi-static noise, and Ornstein-Uhlenbeck noise.

Charge-noise processes in semiconductor materials are generally associated with a noise spectral density that follows a $1/f$ frequency dependence. White noise, such as Johnson-Nyquist noise \cite{Johnson1928,nyquistalt} may also be present, which is observed at lower levels then $1/f$ noise \cite{guidostutorial}. It originates from thermally fluctuating currents in metallic structures, such as metallic gates, which are a major source of Johnson noise  \cite{johnsonnyquistquelle1,johnsonnyquistquelle2}. 

We therefore start our noise analysis with the case of white noise. Its mathematical description is particularly straightforward. This is due to the fact that its noise spectral density is frequency-independent by definition, $S_{kk'}(\omega)=S_{kk'}=\text{const.}$, which applied to Eqs.~\eqref{chi1_k_chikp_tp_von_s} and \eqref{chi_1_quadrat_t_von_s}, leads to the following expressions for the auto- or cross-correlations for different times and for the same time point,
\begin{equation}
    \label{auto_corrrelations_wn}
    \begin{aligned}
    \langle\langle \chi_k(t')\chi_{k'}(t)\rangle\rangle_{\text{w}} &= 2 \pi S_{kk'}\text{min}(t,t'), \\
     \langle\langle \chi_k^2(t)\rangle\rangle_{\text{w}}&=2\pi S_kt.
    \end{aligned}
\end{equation}
The noise correlation spectral density of white noise is frequency-independent, meaning that the correlator for white noise has to be a delta-function in the time domain, since these two quantities are related via Fourier transformation, Eq.~\eqref{auto_and_corr_s_w_def}. The property of white noise that correlations only exist for the same time point \cite{unkorrwnquelle1,unkorrwnquelle2} can be written as
\begin{equation}
    \label{formel_unkorr_wn}
    \langle\langle\delta \omega_{1}(t_1)\delta\omega_{2}(t_2)\rangle\rangle_{\text{w}} = W_{12}\delta(t_1-t_2).
\end{equation}

In contrast, quasi-static noise changes very slowly over the measurement process.  It is important to note that the noise is still a random variable, which means that for each measurement of the cavity emission, the noise has a different outcome, remaining almost constant over the process of one measurement \cite{characterizeqs,qubitdephasingqs}. 

Since the autocorrelation function for quasi-static noise is time independent, the noise spectral density for quasi-static noise must be a delta function in frequency $S_{kk'}(\omega)=S_{kk'}\delta(\omega)$.
For quasi-static noise, we find,
\begin{equation}
    \label{qs_noise_relations}
    \langle\langle\chi_k(t')\chi_{k'}(t)\rangle\rangle_{\text{qs}}=tt'S_{kk'}, \quad \langle\langle\chi_k^2(t)\rangle\rangle_{\text{qs}}=t^2S_k,
\end{equation}
where we consider the case of $t'\leq t$, which will be used in the evaluation of the cavity emission. Note that since the noise spectral density is a delta-function in the case of quasi-static noise, the memory effects of the bath are not washed out. Therefore, the bath typically has strong non-Markovian effects \cite{qubitasprobeqs}. Note that the method presented in \cite{characterizeqs} can be used to extract the noise autocorrelation spectral densities in the case of quasi-static noise.

Noise sources which follow Ornstein-Uhlenbeck (OU) statistics are described by an  exponentially decaying auto-correlator in the time domain, $W_{kk'}(t)=e^{-|t|/\tau_{kk'}}$, where $\tau_{kk}=\tau_{k}$, denotes the auto-correlation (cross-correlation) time constant of the noise  \cite{originaloupaper,machlup1954randomtelegraph,oulowpass}. The Fourier transform of this correlator yields the noise spectral density, 
\begin{equation}
    \label{auto_corr_spec_den_ou}
    S_{kk'}(\omega)=\int_{0}^\infty dt W_{kk'}(t)\frac{\text{cos}(\omega t)}{\pi}=\frac{\Gamma_{kk'}}{\pi(\omega^2+\Gamma_{kk'}^2)},
\end{equation}
where $\Gamma_{kk'} = \tau_{kk'}^{-1}$ is the decay rate of the noise auto-correlation ($k=k'$) or cross-correlation ($k\ne k'$). Applying this formula to Eqs.~\eqref{chi1_k_chikp_tp_von_s} and \eqref{chi_1_quadrat_t_von_s} leads to the noise correlation terms appearing inside the cavity emission, which can be written as 
\begin{equation}
    \begin{aligned}
    \label{chi_relations_ou}
    \langle\langle\chi_k(t')\chi_{k'}(t)\rangle\rangle_{\text{ou}}&=\frac{1 }{\Gamma_{kk'}^2}\bigg(2\Gamma_{kk'}t'-e^{\Gamma_{kk'}(t'-t)}\\
    &\quad +e^{-\Gamma_{kk'}t}+e^{-\Gamma_{kk'}t'}-1\bigg),\\
    \langle\langle\chi_k^2(t)\rangle\rangle_{\text{ou}}&=\frac{2}{\Gamma_k^2}\left( \Gamma_kt-1+e^{-\Gamma_kt}\right).
    \end{aligned}
\end{equation}
Similarly to the quasi-static noise case we are considering the case of $t'\leq t$.

We find that OU noise is an intermediate case between white noise and quasi-static noise: In the limit of high correlation decay rates, we recover the case of white noise. It should be noted that, in order to achieve this, it is necessary to carry out a renormalization by multiplying the noise spectral density by $\Gamma_{kk'}$ prior to taking the limits. Conversely, in the limit of low correlation decay rates, we recover the case of quasi-static noise.


\section{Cavity emission}
\label{sec:cavityemission}
The dynamics of the system can be described by applying Eq.~\eqref{quantum_master_equation} to the operators $\hat{\sigma}_{-k}$ and $\hat{\sigma}_{zk}$, which describe the qubit state, as well as the annihilation operator $\hat{a}$ of photons in the cavity and the frequency-dependent annihilation operator $\hat{b}(\omega)$ of photons at the cavity ports. The resulting system of differential equations is known as the quantum Langevin equations (QLE). Following the general procedure of input-output theory, the differential equation for $\langle\hat{b}(\omega)\rangle$ can be solved assuming a frequency-independent coupling between the cavity and its surroundings, $\kappa(\omega)=\kappa$. Its solution is substituted into the differential equation for $\langle\hat{a}\rangle$. With this, the QLE can be stated as \cite{inoutgardiner}
 \begin{subequations}
 \label{final_set_of_qleqs}
     \begin{align}
         \frac{d\langle \hat{\sigma}_{-k}\rangle}{dt}&=\left[ -i\left(\Delta_k+\delta\omega_{k}(t)\right)-\frac{\gamma_{dk}}{2}\right]\langle\hat{\sigma}_{-k}\rangle 
         +ig_k\langle\hat{\sigma}_{zk}\rangle\langle\hat{a}\rangle, \label{20a}\\
         \frac{d\langle\hat{a}\rangle}{dt}&=\left(-i\Delta_c-\frac{\kappa}{2}\right)\langle\hat{a}\rangle-i\sum_{k=1,2}g_k\langle\hat{\sigma}_{-k}\rangle,  \label{20b}\\
         \frac{d\langle\hat{\sigma}_{zk}\rangle }{dt}&=-\gamma_k-\gamma_k\langle\hat{\sigma}_{zk}\rangle 
         +2ig_k\left(\langle\hat{a}^\dagger\rangle\langle\hat{\sigma}_{-k}\rangle-\langle\hat{a}\rangle\langle\hat{\sigma}_{+k}\rangle\right),\label{20c}
     \end{align}
 \end{subequations}
 in the low temperature limit, $T\ll \Delta\omega_k$. In this limit, stimulated emission and absorption processes can be neglected, whereas at higher temperatures, both processes have to be taken into account. We assume that there is no input field present at both cavity ports. 
 In the low-temperature limit the total noise-independent (Markovian) decoherence rate is defined as $\gamma_{dk}=\gamma_k+2\gamma_{\phi k}$.
 The total cavity loss rate is given by the sum of the loss rates at both cavity ports, $\kappa = \kappa_1+\kappa_2$. For further details on the derivation and simplification of the QLE see Appendix~\ref{sec:appqle}.
 
The quantity of interest in our calculations is the emission from the cavity, which is captured in the output field at the second port of the cavity \cite{inoutgardiner}, 
\begin{equation}
    \label{output_field}
    B_{\text{out}}(t)=-\sqrt{\kappa_2}\langle\hat{a}\rangle. 
\end{equation}
This quantity can be obtained by means of input-output theory \cite{guidostutorialvon2020, inoutgardiner}. The output from the first cavity port could be calculated similarly, but is ignored for the purpose of this study. Solving the QLE exactly is challenging for this system because of the coupling terms $\sim g_k$, which would require considering an infinite number of terms with increasing powers of $g_k$. An approximate solution of the QLE can be found by employing perturbation theory in $g_k$. The condition that justifies this is that the influence of the coupling terms has to be much smaller than the influence of the remaining dynamics (excluding the noise influence), $g_k\ll |\Delta_k|, |\Delta_c|$ and $\gamma_{dk}\ll |\Delta_k|$. 

From this perturbation theory in $g_k$, the solution of the QLE can be used to derive the output field. In order to adequately describe the effect of the noise correlations, it is necessary to take into account contributions up to the third order in $g_k$. The QLE are then separated into different orders in $g_k$ and can be solved step by step, starting with the zeroth order in $g_k$. The solution of the QLE depends on the initial conditions $\langle \hat{\mathcal{O}}_0\rangle= \langle \hat{\mathcal{O}}(t=0)\rangle$ of the system. An initially empty cavity,  $\langle\hat{a}_0\rangle=0 $, is utilized to minimize the complexity of the calculations whilst preserving the impact of noise correlations. The initial states of the qubits can be chosen almost arbitrarily. The only requirement is a non-vanishing initial coherence $\langle \hat{\sigma}_{-k,0}\rangle \neq0$ for at least one of the qubits, lest we find a vanishing cavity emission. The output field can now be calculated with perturbation theory up to the third order in $g_k$ as,
 \begin{equation}
    \label{output_field_term}
     \begin{aligned}
     &B_{\text{out}}(t) =-\sqrt{\kappa_2}e^{-ct}\Bigg( 
     \sum_{k=1,2}g_k^2 \int_0^t dt' e^{-s_kt'}
      \\
     &\quad \times e^{-i\lambda_k\chi_k(t')}\int_0^{t'} dt''
     \big((\langle\hat{\sigma}_{zk,0}\rangle+1)e^{-\gamma_kt''}-1\big)\\
     &\quad\quad \times e^{s_kt''}\mu(t'')e^{i\lambda_k\chi_k(t'')}+\mu(t)\Bigg),
    \end{aligned}
\end{equation}
where we have introduced
 \begin{equation}
    \label{mu_von_t}
     \mu(t)= \sum_{k'=1,2}\frac{g_{k'}}{i}\langle\hat{\sigma}_{-k',0}\rangle\int_0^{t}dt'  e^{-s_{k'}t'}e^{-i\lambda_{k'}\chi_{k'}(t')},
 \end{equation}
 $c= i\Delta_c+\frac{\kappa}{2}$, $q_k= i\Delta_k+\frac{\gamma_{dk}}{2}$, and $s_k= q_k-c$.

The cavity emission captures the effects of the various interaction and photon exchange processes between the cavity and the qubits. The different orders in the coupling term $g_k$ correspond to different physical processes, with higher orders involving more complex processes in which photons are exchanged consecutively. In general, higher order processes are less probable. The first order $\sim g_k$ contributions to the cavity emission consist of the sum of the noise influences from the photons, which were emitted from the two qubits. In even orders ($\sim g_k^2,\sim g_k^4,\dots$) there are no contributions to the cavity emission. This is because the cavity is initially empty and because there is no direct interaction between the qubits.

\begin{figure}
    \centering
    \makebox[\linewidth][c]{%
    \includegraphics[width=0.8\linewidth]{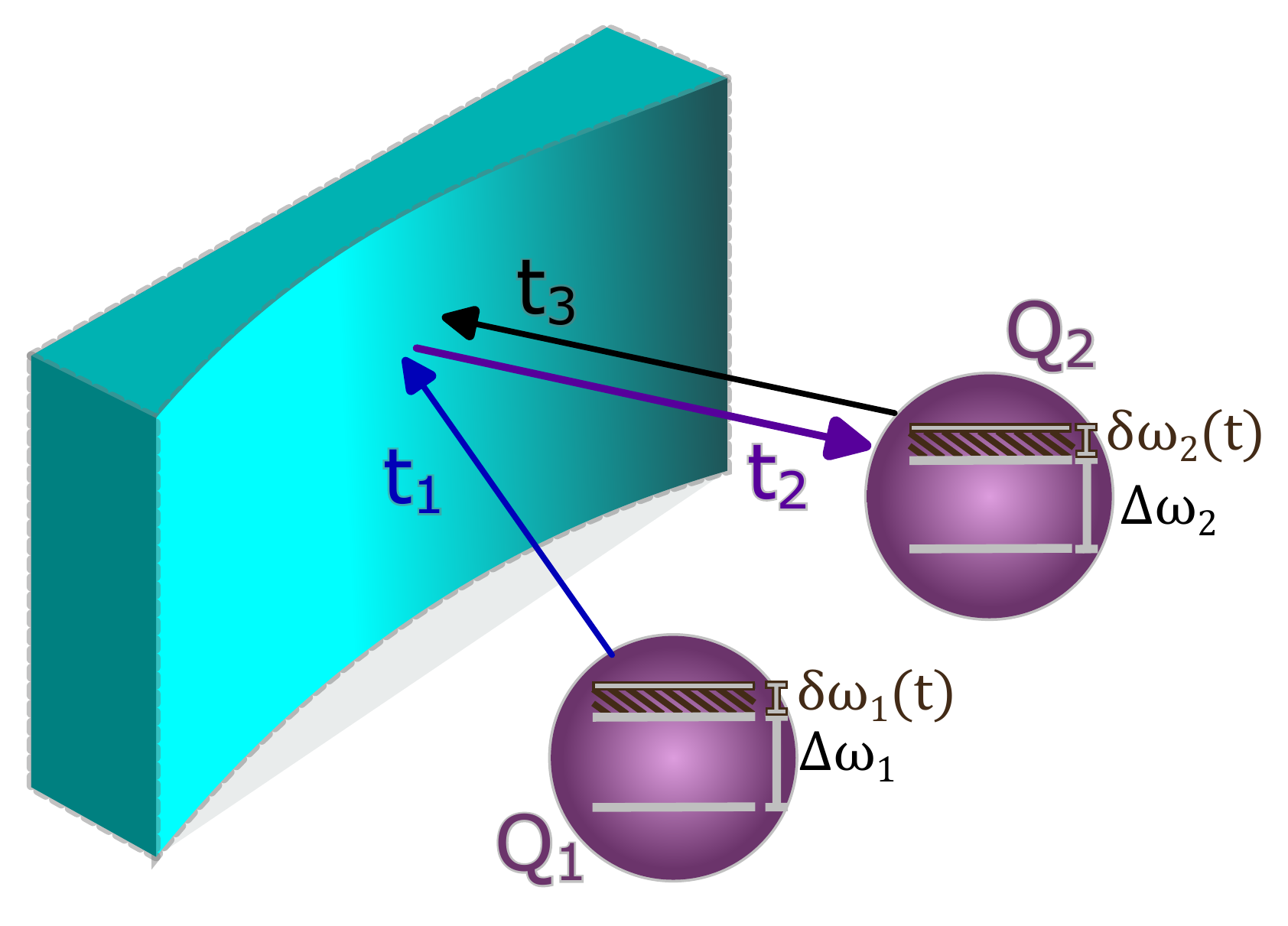}(a)}
\hfill 
    \medskip
    \centering
   \makebox[\linewidth][c]{%
    \includegraphics[width=0.8\linewidth]{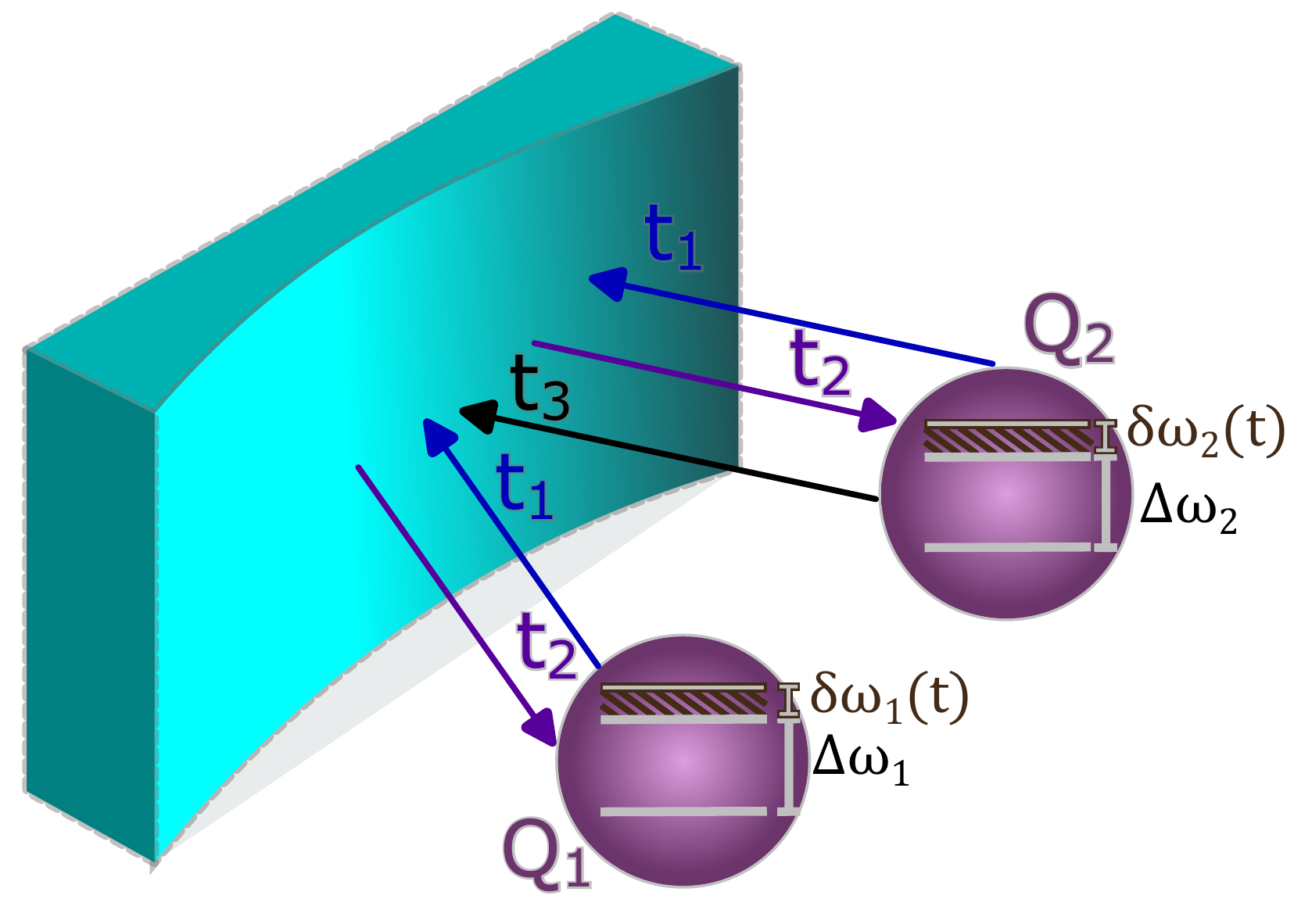}(b)}
\hfill

    \caption{(a) Example illustrating consecutive third-order photon-exchange processes ($\sim g_k^3$) between the qubits and the cavity, consequently resulting in noise correlations in the output of the cavity. At time $t_1$ a photon, which is affected by noise in Q1, is emitted from Q1 into the cavity and absorbed at time $t_2$ by Q2. Finally, at time $t_3$ a photon is emitted by Q2 into the cavity, accumulating the noise effects from both Q1 and Q2, such that the cavity output contains noise cross-correlations.
    (b) Example of consecutive fifth-order ($\sim g_k^5$) photon-exchange processes between the qubits and the cavity. This leads to further noise correlations effects in the output of the cavity, which are strongly suppressed compared with the third order contributions. 
     As photon exchanges are occurring between the cavity and both qubits simultaneously, noise influences from both qubits will also be present at the same time, leading to noise correlations, even in the case of white noise. 
     \label{dritte_und_fuenfte_ordnung_skizze}}
\end{figure}

In third order $\sim g_k^3$, contributions from the noise correlations appear. The first qubit (Q1) emits a photon to which the noise on that qubit contributes into the cavity which is reabsorbed by the second qubit (Q2), and leads to an excitation that is influenced by the noise affecting Q2. Afterwards, a photon is emitted from Q2 into the cavity (see Fig.~\ref{dritte_und_fuenfte_ordnung_skizze}a). Note that this process can also start with an emission of a photon from Q2, which can be described equivalently, simply by exchanging the roles of Q1 and Q2. Figure~\ref{dritte_und_fuenfte_ordnung_skizze}b sketches the possibility of consecutive fifth-order interaction processes between the qubits and the cavity, which are strongly suppressed compared to the third-order processes but allow for some interaction steps to occur at the same time, such as the first emission from both qubits. Remarkably, there is no contribution to the third order $\sim g_k^3$ in the output field of the cavity within the limit of vanishing relaxation rates $\gamma_k\rightarrow 0$. Consequently, in this limit, noise correlations are not detectable. Therefore, we proceed with a finite qubit relaxation rate in our calculations.

The remaining time integrals in the output field expression Eq.~\eqref{output_field_term} can be solved if the time-dependent noise functions $\chi_k(t)$ are known. Since we are striving to determine an unknown noise spectral density, this is not possible in general. Furthermore, it is important to note that these noise functions should differ in each repetition of the measurement since they are random functions. To find an analytic expression for the expected output field, we average over many measurements, which is indicated with $\langle\langle \dots \rangle\rangle$. For Gaussian noise with zero mean, we can express the average of the exponentiated random variable via its second cumulant,
\begin{equation}
    \label{averaging_over_many_measurements_relation_1}
    \langle\langle e^{\pm i\lambda_k\chi_k}\rangle\rangle=e^{-\frac{\lambda_k^2}{2}\langle\langle \chi_k^2\rangle\rangle}.
\end{equation}
After the averaging process over many measurements, the output field can be separated into terms, containing autocorrelations (``ac'') and into terms, containing cross-correlations (``cc''), 
\begin{equation}
    \label{separation_of_output_field}
    \langle\langle B_{\text{out}}(t)\rangle\rangle=\langle\langle B_{\text{out}}^{(\text{ac})}(t)\rangle\rangle+\langle\langle B_\text{out}^{\text{(c)}}(t)\rangle\rangle.
\end{equation}
If the setup is reduced to coupling to only one of the qubits, the output field reduces to the autocorrelation term for the first or second qubit, respectively. In this case, there is no contribution from noise cross-correlations. Therefore, for differentiating between noise cross-correlation effects and autocorrelations, it is possible to calculate the output field for the situation with both qubits and to subtract the emission in the case where only the first qubit is coupled to the cavity, as well as the emission in the case where only the second qubit is coupled to the cavity. The remaining contribution will then give the noise cross-correlations. The autocorrelation terms are given by
\begin{equation}
    \label{auto_correlations}
    \begin{aligned}
    &\langle\langle B_{\text{out}}^{(\text{ac})}(t)\rangle\rangle=\sqrt{\kappa_2}e^{-ct}\bigg( -\mu^{(\text{ac})}(t)+\sum_{k=1,2}\int_0^{t}dt'\\
    & \times ig_k^3 e^{-s_kt'}\int_0^{t'}dt''\big(\left( \langle \hat{\sigma}_{zk,0}\rangle+1\right)e^{-\gamma_kt''}-1\big)e^{s_kt''}\\
    & \times \langle\hat{\sigma}_{-k,0}\rangle\int_0^{t''}dt'''e^{-s_kt'''} e^{-\frac{1}{2}\langle\langle\chi_{\text{(ac),k}}^{2} (t',t'',t''')\rangle\rangle}\bigg),
    \end{aligned}
\end{equation}
where 
\begin{subequations}
    \label{chi_k_quadrat_abkuerzung}
    \begin{align}
        &\mu^{(\text{ac})}(t)=-i\sum_{k'=1,2}g_{k'}\langle\hat{\sigma}_{-k',0}\rangle\int_0^t dt' e^{-s_{k'}t'}\notag \\
        &\quad \times e^{-\frac{\lambda_{k'}^2}{2}\langle\langle \chi_{k'}^2(t')\rangle\rangle},\\
        &\langle\langle \chi_{\text{(ac),k}}^2(t',t'',t''')\rangle\rangle=\lambda_k^2\bigg(\langle\langle\chi_k^2(t')\rangle\rangle+\langle\langle\chi_k^2(t'')\rangle\rangle \notag \\
        &\quad +\langle\langle \chi_k^2(t''')\rangle\rangle+2\langle\langle\chi_k(t')\chi_k(t''')\rangle\rangle \notag\\
        &\quad -2\langle\langle\chi_k(t')\chi_k(t'')\rangle\rangle
         -2\langle\langle\chi_k(t'')\chi_k(t''')\rangle\rangle\bigg).
    \end{align}
\end{subequations}
The averaged noise cross-correlation terms can be written as 
\begin{equation}
    \label{correlations_average_over_many_measurements}
    \begin{aligned}
        &\langle\langle B_{\text{out}}^{(\text{cc})}(t)\rangle\rangle=\sqrt{\kappa_2}e^{-ct} \sum_{k=1,2}\sum_{\substack{ k'=1,2\\k'\ne k}}\int_0^{t}dt'ig_k^2e^{-s_kt'}\\
        & \times g_{k'}\int_0^{t'}dt''\bigg(-1+\left(\langle\hat{\sigma}_{zk,0}\rangle+1\right)e^{-\gamma_kt''}\bigg)e^{s_kt''} \\
       & \times \langle\hat{\sigma}_{-k',0}\rangle \int_0^{t''} dt'''e^{-s_k't'''}e^{-\frac{1}{2}\langle\langle\chi_{\text{(cc),kk'}}^{2}(t',t'',t''')\rangle\rangle},
    \end{aligned}
\end{equation}
where 
\begin{equation}
    \label{chi1_chi2_abkuerzung}
    \begin{aligned}
        &\langle\langle \chi_{\text{(cc),kk'}}^2(t',t'',t''')\rangle\rangle=\lambda_k^2\big( \langle\langle\chi_k^2(t')\rangle\rangle+\langle \langle \chi_k^2(t'')\rangle\rangle\\
        &\ -2\langle\langle\chi_k(t')\chi_k(t'')\rangle\rangle\big)+2\lambda_k\lambda_{k'}\big( \langle\langle\chi_k(t')\chi_{k'}(t''')\rangle\rangle\\
        &\ -\langle\langle\chi_k(t'')\chi_{k'}(t''')\rangle\rangle\big)+\lambda_{k'}^2\langle\langle\chi_{k'}^2(t''')\rangle\rangle.
    \end{aligned}
\end{equation}
In the following sections of this paper, the output field of the cavity will be analyzed in more detail. This will be achieved by calculating the auto- and cross-correlation terms for the cases of white noise, quasi-static noise, and Ornstein-Uhlenbeck noise. For these processes, the auto- and cross-correlators can be related  to the noise spectral densities $S_{k}(\omega)$ and the noise correlation spectral density $S_{12}(\omega)$, such that the time integrals and the frequency integral can be solved.

The design of an experiment for extracting $S_{12}(\omega)$ is described below. Three measurements have to be performed: the first and second measurements are realized for the situation, where only Q1 or Q2 are coupled to the cavity. The cavity emission of these experiments is denoted by $\langle\langle \tilde{B}_{\text{out}}^{q1}\rangle\rangle$ and $\langle\langle \tilde{B}_{\text{out}}^{q2}\rangle\rangle$, respectively. Note that $\sim$ is used to indicate that the quantity is measured. From these measurements, the respective noise spectral densities $S_k(\omega)$ have to be extracted. A method for extracting the noise spectral density is presented in \cite{pm}. In the third measurement, both qubits are coupled to the cavity. This result is designated with $\langle\langle\tilde{B}_{\text{out}}(t)\rangle\rangle$. If then the cavity emission contingents of the individual qubits are subtracted from the cavity emission, there will be three remaining parts.
The first part consists of the noise-free output, $B_{\text{out}}^{\text{q1,q2}}(t)$, which must be calculated from Eq.~\eqref{no_noise_correlation_termssymmetric}. The second part is given by an additional contribution to the auto-correlation terms, $\langle\langle B_{\text{out}}^{\text{qk+}}(t)\rangle\rangle$, which is only present if both qubits are coupled to the cavity. This quantity has to be calculated as well, and will be shown for several types of noise in the following sections. Finally, the third part is the desired correction term for the noise correlations, $\langle\langle B_{\text{out}}^{\text{q1,q2}}(t)\rangle\rangle$. The extraction of the noise correlation contribution is performed by using
\begin{equation}
    \label{noise_corr_reconstruction_from_experiments_eq}
    \begin{aligned}
    &\langle\langle B_{\text{out}}^{\text{q1,q2}}(t)\rangle\rangle=\langle\langle\tilde{B}_{\text{out}}(t)\rangle\rangle-\langle\langle \tilde{B}_{\text{out}}^{q_1}(t)\rangle\rangle\\
    &-\langle\langle \tilde{B}_{\text{out}}^{q_2}(t)\rangle\rangle-B_{\text{out}}^{\text{q1,q2}}(t)-\langle\langle B_{\text{out}}^{\text{qk+}}(t)\rangle\rangle.
    \end{aligned}
\end{equation}

In order to fully observe the influence of the noise, the noise-free case is examined in more detail. In this situation, it is possible to solve the time integrals analytically. Also, it is beneficial to separate further between terms which are influenced by only one of the qubits, $B_{\text{out}}^{q_k}(t)$, and terms which capture influences from both qubits, $B_{\text{out}}^{\text{q1,q2}}(t)$. The total output field is the sum of these two terms, $B_{\text{out}}(t)\ = \ B_{\text{out}}^{q_k}(t)+B_{\text{out}}^{\text{q1,q2}}(t)$. The contributions of the single qubits can be calculated as
\begin{equation}
    \label{noise_free_single_qubits}
    \begin{aligned}
   & B_{\text{out}}^{q_k}(t)=i\sqrt{\kappa_2}\sum_{k=1,2}-\frac{\langle\hat{\sigma}_{-k,0}\rangle}{s_k}\bigg[g_k\big( e^{-q_kt}-e^{-ct}\big)\\
    &\quad +\frac{g_k^3}{s_k}\bigg( t\left(e^{-ct}+e^{-q_kt}\right)-\frac{2(e^{-ct}-e^{-q_kt})}{s_k}\bigg)\\
    &\quad +g_k^3 \big(\langle\hat{\sigma}_{zk,0}\rangle+1\big)\bigg[ \frac{e^{-(\gamma_k+q_k)t}-e^{-ct}}{\gamma_ks_{k+}}\\
    &\quad+\frac{e^{-(c+\gamma_k)t}-e^{-q_kt}}{\gamma_ks_{k-}}\bigg]\bigg],
    \end{aligned}
\end{equation}
where we have introduced the new variables  $s_{k\pm}= q_k-c\pm \gamma_k$. In the fully symmetric case, which refers to  $g_1=g_2= g, \quad \langle \hat{\sigma}_{-1,0}\rangle=\langle\hat{\sigma}_{-2,0}\rangle=\langle\hat{\sigma}_{-,0}\rangle, \langle\hat{\sigma}_{z1,0}\rangle=\langle\hat{\sigma}_{z2,0}\rangle= \langle\hat{\sigma}_{z,0}\rangle, \quad \Delta_1=\Delta_2= \Delta ,
        \gamma_{\phi 1}=\gamma_{\phi 2}=\gamma_\phi, \gamma_1=\gamma_2= \gamma,\rightarrow \gamma_{d1}=\gamma_{d2}=\gamma_d, \rightarrow q_1=q_2= q, s_1=s_2= s,
        \rightarrow s_{1\pm}=s_{2\pm}= s_{\pm}$, the correlations between both qubits are given by
\begin{equation}
    \label{no_noise_correlation_termssymmetric}
    \begin{aligned}
        &B_{\text{out}}^{\text{q1,q2}}(t) \ = \  -\frac{2ig_k^3\sqrt{\kappa_2}\langle\hat{\sigma}_{-,0}\rangle}{s}\bigg[\frac{2}{s^2}\big(e^{-qt}-e^{-ct}\big)\\
        &+\frac{t\big( e^{-ct}+e^{-qt}\big)}{s}+\big(\langle\hat{\sigma}_{z,0}\rangle+1\big)\bigg[\frac{e^{-(q+\gamma)t}
        -e^{-ct}}{s_{+}\gamma} \\
        &+\frac{e^{-(c+\gamma)t}-e^{-qt}}{s_{-}\gamma} \bigg]\bigg].
    \end{aligned}
\end{equation}

\section{White noise}
\label{sec:white}
Here, we extend the study of the impact of noise auto-correlations on cavity emission in the context of white noise \cite{pm} to the cross-correlations. After plugging in the relations for white noise, Eq.~\eqref{auto_corrrelations_wn} into the averaged output field, Eqs.~\eqref{auto_correlations} and \eqref{correlations_average_over_many_measurements}, an analytic expression for the output field in the presence of white noise can be obtained. We find that up until the third order in $g_k$, all noise correlation terms cancel out and the output field is independent of $S_{12}(\omega)$. This can be observed from the cross-correlator term inside the output field Eq.~\eqref{chi1_chi2_abkuerzung}, which reads
\begin{equation}
    \label{no_noise_corr_for_white_noise}
    \langle\langle\chi_{\text{(cc),kk'}}^{2}(t',t'',t''')\rangle\rangle_{\text{w}}=2\pi \big[ \lambda_{k}^2 S_k \big(t'-t''\big)+\lambda_{k'}^2S_{k'}t'''\big],
\end{equation}
for white noise and which is independent of $S_{12}$. For this relation, we used the fact that the noise cross-correlation terms always appear inside the time-ordered integrals where $t\geq t'\geq t''\geq t'''$ inside the cavity emission. This observation corresponds to the fact that multivariate distributed Gaussian white noise can only show correlations between the noise on the different qubits for the same time. For different times, white noise does not show correlations \cite{unkorrwnquelle1,unkorrwnquelle2}. For an intuitive explanation for the absence of the correlation effects up to third order in $g_k$, we refer to Fig.~\ref{dritte_und_fuenfte_ordnung_skizze}a. Since for fifth-order photon exchange processes photon emission from Q1 and Q2 followed by an absorption from the cavity can happen at the same time, cross-correlation effects appear, see Fig.~\ref{dritte_und_fuenfte_ordnung_skizze}b.

\section{Quasistatic noise}
\label{sec:quasi}
Next, we extend the known result for the cavity output in the presence of quasi-static noise \cite{pm} to include cross-correlations. Using Eqs.~\eqref{separation_of_output_field}, \eqref{auto_correlations}, and  \eqref{correlations_average_over_many_measurements}, we find
\begin{subequations}
    \label{relations_quasistatic_for_output_field}
    \begin{align}
        \mu^{(\text{ac})}(t)\ =-i&\sum_{k'=1,2}g_{k'} \langle \hat{\sigma}_{-k',0}\rangle\\
         &\times 
         \int_0^t dt' e^{-q_{k'}t'}
        e^{-\frac{\lambda_{k'}^2}{2}S_{k'}(t')^2},\\
        \langle\langle\chi_{\text{(ac),k}}^2 (t',t'',t''')\rangle\rangle_{\text{qs}}\ &= \lambda_k^2S_k(t'-t''+t''')^2,\\
        \langle\langle\chi_\text{(cc),kk'}^2(t',t'',t''')\rangle\rangle_{\text{qs}}\ &= \lambda_k^2S_k(t'-t'')^2\notag \\
        \quad +2\lambda_k\lambda_{k'}S_{12}t'''&(t'-t'')+\lambda_{k'}^2S_{k'}(t''')^2.
    \end{align}
\end{subequations}
These integrals can be calculated numerically or by way of  a Taylor series expansion in the noise sensitivity, which we assume to be equal for both qubits for simplicity here, $\lambda_1=\lambda_2= \lambda$.  Up to the second order in $\lambda$, we find 
$e^{-\frac{\lambda^2}{2}\langle\langle\dots\rangle\rangle}\approx1-\lambda^2\langle\langle\dots\rangle\rangle /2$.
This approximation is valid if $\lambda^2\langle\langle\dots\rangle\rangle /2 \ll 1$.  For the noise correlation terms $B_{\text{out}}^{\text{q1,q2}}(t)+\langle\langle B_{\text{out}}^{\text{q1,q2}}(t)\rangle\rangle_{qs}$, this inequality indicates that the approximation is only valid for the beginning of the time evolution, 
\begin{equation}
    \label{condition_for_tse_in_lambda_noi_corr}
    t\ll \frac{\sqrt{2}}{\lambda\sqrt{S_1+S_2+2S_{12}}}.
\end{equation}
 
The noise correlation contribution is obtained by subtracting the individual qubit contributions, the noise-free interaction part, and the additional autocorrelation contribution from the total cavity emission according to Eq.~\eqref{noise_corr_reconstruction_from_experiments_eq}. Note that we consider the fully symmetric situation, where we additionally assume equal noise spectral densities $S_1=S_2= S_q$ and equal noise sensitivities, $\lambda_1=\lambda_2= \lambda$. The cavity emission from the experiments, where only one qubit is coupled to the cavity, can be separated into a noise-free interaction part and a noise correction part, which describes the autocorrelation effects, $\langle\langle \tilde{B}_{\text{out}}^{\text{qk}}(t)\rangle\rangle=B_{\text{out}}^{\text{qk}}(t)+\langle\langle B_{\text{out}}^{\text{qk}}(t)\rangle\rangle$. The noise-free part can be subtracted from the measured quantity by applying Eq.~\eqref{noise_free_single_qubits}. The remaining part can be solved for the noise spectral density $S_{q}$. It is given by
\begin{equation}
    \label{output_field_for_qs_noise_in_tse}
    \begin{aligned}
        \langle\langle B_\text{out}^{qk}(t)\rangle\rangle_{\text{qs}}&=-\frac{i\langle\hat{\sigma}_{-,0}\rangle g S_q \sqrt{\kappa_2}\lambda^2e^{-ct}}{2s^3}\\
        &\quad \times \bigg( 2 +e^{-st}(-2-st(2+st)) \bigg),
    \end{aligned}
\end{equation}
where we have linearized the equation in the cavity-qubit coupling $g$. The full expression until third order in $g$ can be found in Eq.~\eqref{output_field_for_qs_noise_in_tse_APP}. Recovering the noise correlation contribution requires the subtraction of the individual qubit contributions measured in the separate experiments, as well as the noise-free interaction terms (see Eq.~\eqref{no_noise_correlation_termssymmetric}) and from the additional autocorrelation contribution. The latter can be calculated from 
\begin{equation}
    \label{b_qk+_add_term_qs}
    \begin{aligned}
    &\langle\langle B_{\text{out}}^{\text{qk+}}(t)\rangle\rangle_{\text{qs}}=\frac{i\langle\hat{\sigma}_{-,0}\rangle g^3S_q \lambda^2 \sqrt{\kappa_2}e^{-(s+c)t} }{s^3}\\
    &\times\bigg(\frac{2st^3}{3}
    -\frac{\langle\hat{\sigma}_{z,0}\rangle+1}{s_{-}^3\gamma^3}s^4\big(4-\frac{2(6+st)\gamma}{s}+\gamma^2t^2\big)\\
    &+\frac{\langle\hat{\sigma}_{z,0}\rangle+1}{s_{+}^3\gamma^3}e^{-\gamma t}s^4\big(4+\frac{2(6+st)\gamma}{s}+\gamma^2t^2\big)\bigg),
    \end{aligned}
\end{equation}
where we neglected the fast decaying exponential functions and used that $|s|\gg \gamma$. The full expression can be found in Eq.~\eqref{b_qk+_add_term_qs_APP}.

\begin{figure}
    \centering
    \includegraphics[width=\columnwidth]{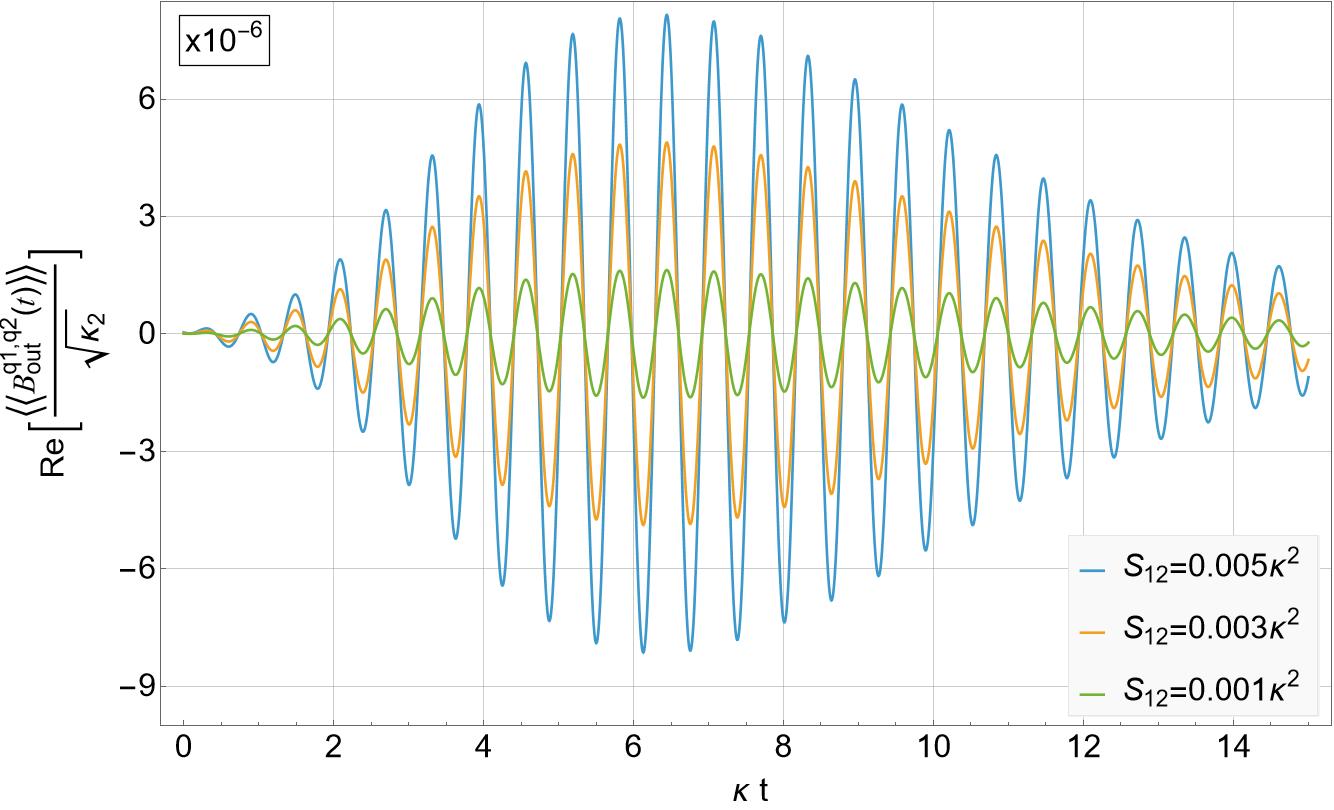}
    \caption{The effect of noise cross-correlations on the time-dependent cavity emission in the presence of quasi-static noise. We plot the contribution of the cavity emission containing the noise cross-correlations, $\langle\langle B_{\text{out}}^{\text{q1,q2}}(t)\rangle\rangle_{\text{qs}}$, from Eq.~\eqref{qs_correlation_terms_qs}, in second order in $\lambda$.
    The parameters are chosen equally for both dots, as $g/(2\pi)=0.5 \ \text{MHz}, \langle\hat{\sigma}_{-,0}\rangle=0.5, \langle\hat{\sigma}_{z,0}\rangle=0, \kappa_2=\frac{\kappa}{2}=0.5 \ \text{MHz}, \lambda=0.9, \gamma_d=1 \ \text{MHz}, \Delta/(2\pi)=-10 \ \text{MHz}, \Delta_c=0$. The condition Eq.~\eqref{condition_for_tse_in_lambda_noi_corr} restricts the validity of the second-order approximation to times  $t\lesssim 10/\kappa  $.}
    \label{qs_bnc_q1_q2_von_s12_plot}
\end{figure}

The remaining contribution recovers the noise correlation term, which can be calculated according to 
\begin{equation}
    \label{qs_correlation_terms_qs}
    \begin{aligned}
        &\langle\langle B_{\text{out} }^{q1,q2}(t)\rangle\rangle_{\text{qs}}=\frac{2i\langle\hat{\sigma}_{-,0}\rangle g^3S_{12}\lambda^2\sqrt{\kappa_2}e^{-ct}}{s^2}
        \\
        &\quad \times \bigg(e^{-st}\bigg(\frac{t^3}{6}+\frac{\langle\hat{\sigma}_{z,0}\rangle+1}{s_{-}^2\gamma^3}s\big(2s-s\gamma t-4\gamma\big)\bigg)\\
        &\quad \quad \quad-\frac{\big(\langle\hat{\sigma}_{z,0}\rangle+1\big)e^{-s_{+}t}s}{s_{+}^2\gamma^3}\big(2s+s_{+}\gamma t+4\gamma\big)\bigg),
    \end{aligned}
\end{equation}
where we again neglect the fast decaying exponential terms and use $|s|\gg \gamma$. The full expression can be found in  Eq.~\eqref{qs_correlation_terms_qs_APP}. Rearranging this equation yields the noise-correlation spectral density $S_{12}(\omega)$. This equation can be applied if the condition Eq.~\eqref{condition_for_tse_in_lambda_noi_corr} is satisfied. The time dependence of the noise correlation part $\langle \langle B_{\text{out}}^{\text{q1,q2}}(t)\rangle\rangle_{\text{qs}}$ for different values of $S_{12}$ derived from Eq.~\eqref{qs_correlation_terms_qs} is illustrated in Fig. ~\ref{qs_bnc_q1_q2_von_s12_plot}. The effects of noise correlations result in a change in the total amplitude of the cavity emission. This amplitude exceeds the noise correlation effects by approximately two orders of magnitude. The noise-correlation dependent part of the cavity emission scales linearly with $S_{12}$ if the noise influence is small compared to the remaining dynamics. The noise correlation effects decay over the same time range as the total cavity emission. The time at which the maximum of $\langle \langle B_{\text{out}}^{\text{q1,q2}}(t)\rangle\rangle_{\text{qs}}$ is reached is independent of $S_{12}$.


\section{Ornstein-Uhlenbeck noise}
\label{sec:ou}
To fill the gap between white noise and quasistatic noise, we now turn to the noise correlation effects resulting from multipartite Gaussian distributed Ornstein-Uhlenbeck (OU) noise. The relations for OU distributed noise (see Eq.~\eqref{chi_relations_ou}) can be substituted into the equations for the output field (Eqs.~\eqref{auto_correlations} and \eqref{correlations_average_over_many_measurements}), leading to the terms containing the autocorrelation noise influence, which are given by
\begin{subequations}    \label{auto_correlation_term_for_ou}
    \begin{align}
         &\mu^{(\text{ac})}(t)=-i\sum_{k'=1,2}g_{k'}\langle\hat{\sigma}_{-k',0}\rangle
        \int_0^t dt' e^{-s_{k'}t'} \notag\\
        &\quad \times\text{exp}\left[\frac{\lambda_{k'}^2}{\Gamma_{k'}^2}\left(1-\Gamma_{k'}t'-e^{-\Gamma_{k'}t'}\right) \right],\\
        &\langle\langle \chi_{\text{(ac),k}}^2(t',t'',t''')\rangle\rangle=\frac{2\lambda_k^2}{\Gamma_k^2}\bigg(-2+\Gamma_k(t'-t''+t''')\notag \\
        &\quad +e^{-\Gamma_kt'''} -e^{-\Gamma_kt''}+e^{-\Gamma_kt'}+e^{\Gamma_k(t''-t')} \notag \\
        &\quad +e^{\Gamma_k(t'''-t'')}-e^{\Gamma_k(t'''-t')}\bigg) ,
    \end{align}
\end{subequations}
while the terms containing the cross-correlation noise influence inside the cavity emission can be written as
\begin{equation}
    \label{correlation_term_for_ou}
    \begin{aligned}
         &\langle\langle \chi_{\text{(cc),kk'}}^2(t',t'',t''')\rangle\rangle_{\text{OU}}=\frac{2\lambda_{k'}^2}{\Gamma_{k'}^2}\bigg(\Gamma_{k'}t'''-1
          +e^{-\Gamma_{k'}t'''}\bigg)\\
         & +\frac{2\lambda_k^2}{\Gamma_k^2}\big( e^{\Gamma_k(t''-t')}-1+\Gamma_k(t'-t'')\big)\\
         &  +\frac{2\lambda_k\lambda_{k'}}{\Gamma_{12}^2}\left( e^{-\Gamma_{12} t''}
         -e^{-\Gamma_{12}t'}\right)\left(e^{\Gamma_{12}t'''}-1\right).
    \end{aligned}
\end{equation}
Note that the first two lines in Eq.~\eqref{correlation_term_for_ou} give the additional contribution to the autocorrelation terms, which emerge exclusively in the presence of both qubits, while the last two lines in this equation refer to the actual noise correlation terms. Given that the cavity emission in the case of OU noise cannot be solved analytically, a Taylor series expansion in $\lambda$, the qubit sensitivity with respect to noise, must be performed again.  

For higher values of the cross-correlation decay rate, $\lambda/\Gamma_{12}\ll1$, the approximation is highly accurate for the entire time evolution. In the opposite situation, where the cross-correlation decay rate is small, $\lambda/\Gamma_{12}\gg1$, the approximation is accurate only for sufficiently short times. In this case, it is beneficial to model the noise as quasi-static instead. With this, we find an analytic expression for the cavity emission. We can separate the output field into different contributions according to Eq.~\eqref{noise_corr_reconstruction_from_experiments_eq}. We consider the fully symmetric situation again, where we additionally assume equal noise autocorrelation decay rates $\Gamma_1=\Gamma_2= \Gamma$ and equal noise sensitivities $\lambda_1=\lambda_2= \lambda$ for both qubits. 

The autocorrelation terms can be determined from measurements when only one qubit is coupled to the cavity, after subtracting the noise-free contribution Eq.~\eqref{noise_free_single_qubits}. Recovering this autocorrelation influence can be achieved from $\langle\langle B_{\text{out}}^{\text{qk}}(t)\rangle\rangle_{\text{OU}}=\langle\langle \tilde{B}_{\text{out}}^{\text{qk}}(t)\rangle\rangle_{\text{OU}}- B_{\text{out}}^{\text{qk}}(t)$. The noise autocorrelation decay rate $\Gamma$ can be obtained by numerically solving 
\begin{equation}
    \label{auto_corr_ou_linearer_term}
    \begin{aligned}
    &\langle\langle B_{\text{out}}^{\text{qk}}(t)\rangle\rangle_{\text{OU}}=\frac{ig\langle\hat{\sigma}_{-,0}\rangle\sqrt{\kappa_2}\lambda^2}{\Gamma^2}\bigg( \frac{e^{-(q+\Gamma)t}-e^{-ct}}{s+\Gamma} \\
    &\quad +\frac{e^{-ct}(s-\Gamma)+e^{-qt}(\Gamma-s+s\Gamma t)}{s^2}\bigg),
    \end{aligned}
\end{equation}
where we have linearized the equation in the qubit-cavity coupling $g$. The more accurate version can be found in Eq.~\eqref{auto_corr_ou_linearer_term_APP}. 
After recovering the noise autocorrelation decay rate $\Gamma$, the additional contribution to the autocorrelations, which is only present if both qubits are coupled to the cavity, can be calculated by applying 
\begin{equation}
    \label{bqk_plus_ou}
    \begin{aligned}
        &\langle\langle B_{\text{out}}^{\text{qk+}}(t)\rangle\rangle_{\text{OU}}=2\sqrt{\kappa_2}e^{-ct}\int_0^{t}dt'ig^3e^{-st'} \\
    &\quad \times \int_0^{t'}dt''\big(-1+\left( \langle \hat{\sigma}_{z,0}\rangle+1\right)e^{-\gamma t''}\big) e^{st''}\langle\hat{\sigma}_{-,0}\rangle \\
    &\quad \times \int_0^{t''}dt'''e^{-st'''} \frac{\lambda^2}{\Gamma^2}\bigg[2-\Gamma(t'''-t''+t')\\
    &\quad -e^{-\Gamma t'''} - e^{\Gamma(t''-t')}\bigg].
    \end{aligned}
\end{equation}
Finally, all remaining contributions can be separated from the total output of the cavity in the case where both qubits are coupled to the cavity according to Eq.~\eqref{noise_corr_reconstruction_from_experiments_eq}. The remaining part gives the correction from the noise correlations, which can be stated as 
\begin{equation}
    \label{noi_corr_for_ou_term}
    \begin{aligned}
    &\langle\langle B_{\text{out}}^{\text{q1,q2}}(t)\rangle\rangle_{\text{OU}}
    =\frac{2ig^3 \langle\hat{\sigma}_{-,0}\rangle \sqrt{\kappa_2}\lambda^2}{s^2 \Gamma_{12}^2}\bigg[  \alpha_{-}e^{-(q+\gamma)t}\\
    &\quad +\bigg(\frac{2}{\Gamma_{12}}+t-\alpha_{-}\bigg)e^{-(q+\Gamma_{12})t}+\alpha_{+}e^{-(q+\gamma+\Gamma_{12})t}\\
    &\quad +\bigg(-\frac{2}{\Gamma_{12}}+t-\alpha_{+}\bigg)e^{-qt}\bigg],\\
    \end{aligned}
\end{equation}
where we introduced  $\alpha_{\pm}=\Gamma_{12}(\langle\hat{\sigma}_{z,0}\rangle+1)/(\gamma(\Gamma_{12}\pm\gamma))$. Additionally, we neglect the fast decaying exponential terms and consider the regime where $|s|\gg \gamma$ and $|s|\gg \Gamma_{12}$. The more general version of this formula, where the limiting cases are not applied, can be found in Eq.~\eqref{noi_corr_for_ou_term_APP}. This equation can be solved numerically for the correlation decay rate $\Gamma_{12}$, allowing the full characterization of the OU type noise cross-correlations.

 The noise correlations lead to an oscillating contribution to the cavity emission, which results in a change of the total cavity emission. The oscillation period of the noise correlation term is determined primarily by the cavity-probe detuning $\Delta_k$. For correlation decay rates of a smaller magnitude, strong noise correlation effects are achieved, while the noise correlations are washed out for larger correlation decay rates. In both cases, oscillations appear over the duration of the cavity emission, which decay to zero. 

\begin{figure}
    \centering
    \includegraphics[width=\columnwidth]{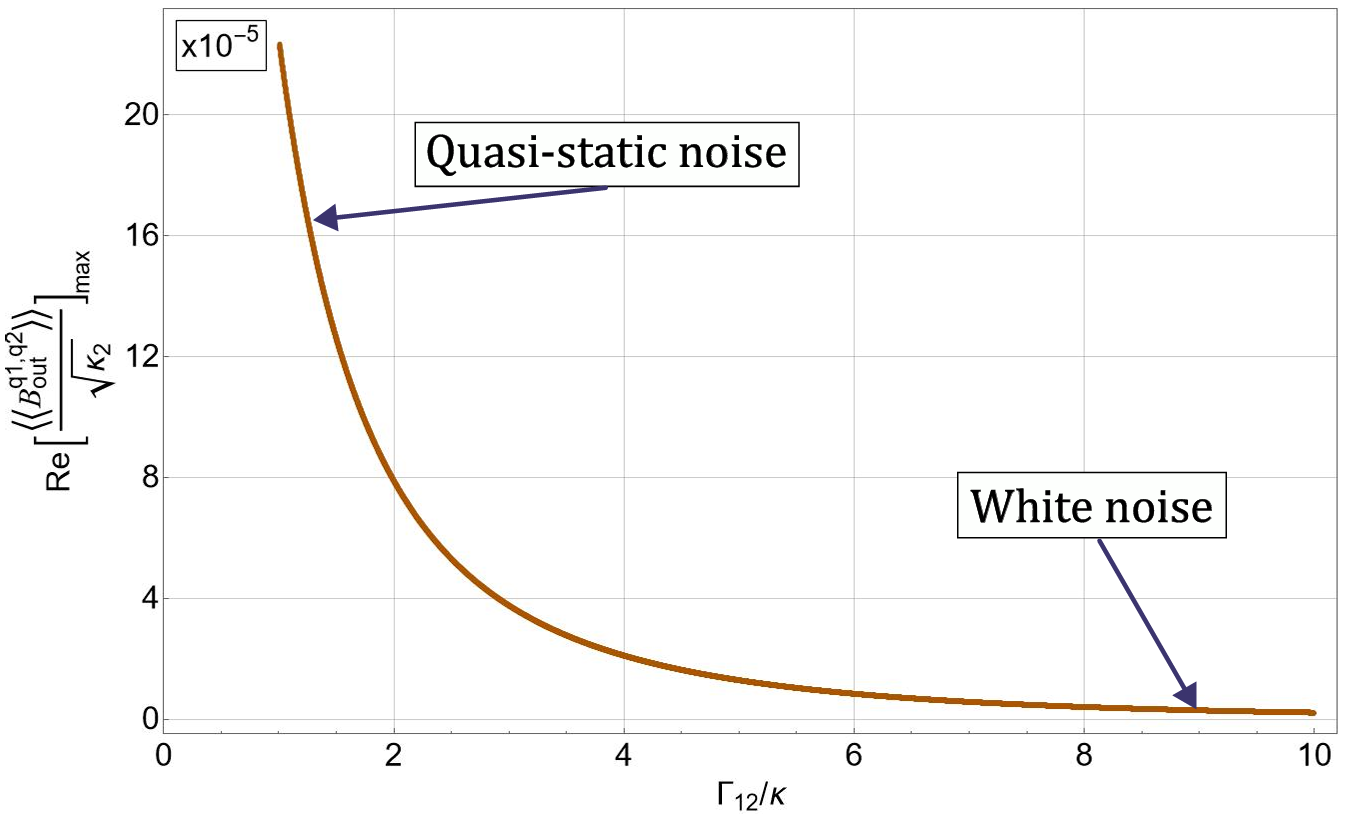}
    \caption{Dependence of the correlation decay rate on the maximum value of $\langle\langle B_{\text{out}}^{\text{q1,q2}}(t)\rangle\rangle_{\text{OU}}$. For low correlation decay rate values, the slope of this curve is approximately linear, which resembles the situation of quasi-static noise. Conversely, if $\Gamma_{12}$ is high, the impact of noise correlation is diminished, and the case of white noise is recovered.
    The parameters are chosen fully symmetric between the two qubits, as $g/(2\pi)=0.5 \ \text{MHz}, \langle\hat{\sigma}_{-,0}\rangle=0.5,\langle\hat{\sigma}_{z,0}\rangle=0,\kappa_2=\frac{\kappa}{2}=0.5 \ \text{MHz},\lambda=0.9,\gamma_d=1 \ \text{MHz}, \Delta/(2\pi)=-10 \ \text{MHz}, \Delta_c=0$. }
    \label{maximum_ou_plot}
\end{figure}

From Eq.~\eqref{noi_corr_for_ou_term}, it is also possible to find the maximal cavity emission in dependence of the correlation decay rate. For this, it is beneficial to move into another rotating frame, by multiplying all oscillating terms $\sim e^{-qt}$ by $ie^{i\Delta t}$, which eliminates the oscillations if $\Delta_c=0$. The correlation decay rate dependence of this maximum value is illustrated in Fig.~\ref{maximum_ou_plot}, where the limiting behaviors for high and low correlation decay rates are clearly visible. For low correlation decay rates, a slope that is approximately linear in $\Gamma_{12}$ is obtained, as in the situation of quasi-static noise. In contrast, the noise correlation contribution is washed out in the limit of high correlation decay rates, as is the case for the presence of white noise. The cases with an intermediate correlation decay rate can be described by OU noise, where $\Gamma_{12}$ can be extracted from Eq.~\eqref{noi_corr_for_ou_term}. The noise correlation terms for OU noise have higher magnitudes compared with quasi-static noise, but in the case of quasi-static noise, the noise correlation terms decay faster (see Fig.~\ref{qs_bnc_q1_q2_von_s12_plot}).

\section{Extracting $S_{12}(\omega)$}
\label{sec:general}
After studying the extraction of the noise correlation spectral density $S_{12}(\omega)$ in the cases of quasi-static noise in Sec.~\ref{sec:quasi} and of OU distributed noise in Sec.~\ref{sec:ou}, we now present a general procedure that enables the extraction of the frequency dependence of the spectral density of the noise cross-correlations. For this procedure, the convolution theorem has to be applied. By extracting the frequency spectrum of the noise correlations, it is also possible to determine the type of noise to which the system is exposed. Therefore, the method presented in this section enables the identification of the dominant sources of noise by determining the spectrum of the noise correlations.

In the following, we consider the case where the sensitivity  to the noise of the qubits is adjustable. In addition, we are now requiring specific initial states for the qubits. In our case, Q1 is initially in a superposition of its ground state and its excited state, while Q2 is initialized in its ground state (interchanging the roles of Q1 and Q2 leads to the same result), 
\begin{equation}
    \label{initial_states_for_conv_theo}
    \begin{aligned}
    |\psi_{Q1}\rangle (0)&=\frac{1}{\sqrt{2}} \left(|\uparrow\rangle+|\downarrow\rangle\right), \\ 
    |\psi _{Q2}\rangle(0)&=|\downarrow\rangle\langle\downarrow|.
    \end{aligned}
\end{equation}
From these initial conditions, we get $\langle\hat{\sigma}_{-1,0}\rangle=\frac{1}{2}$,
$\langle\hat{\sigma}_{-2,0}\rangle=0,\langle\hat{\sigma}_{z1,0}\rangle=0,\langle\hat{\sigma}_{z2,0}\rangle=-1$,
which leads to a significant reduction in the complexity of the equation for the output signal. It is important to note that here we are not considering the fully symmetric case. Now, for the extraction of the noise correlation spectral density, we determine the derivative of the output field with respect to the sensitivity to noise, $\lambda_1$ and $\lambda_2$. When the relation for the correlator given in Eq.~\eqref{chi1_k_chikp_tp_von_s} is substituted into the equation for the output field, after it is averaged over many measurements, see Eqs.~\eqref{separation_of_output_field}, \eqref{auto_correlations} and \eqref{correlations_average_over_many_measurements}, the resulting time integrals can be solved. Note that we are assuming a symmetric and real spectrum for the noise correlation spectral density, $S_{12}(-\omega)=S_{12}(\omega)$. This is generally the case in the presence of classical noise. The resulting integral over the angular frequency contains a time-dependent and a time-independent part. For simplification, we neglect the time-dependent contribution to the angular momentum integral. The error that appears from this will be estimated in the end of this section. In the end, we can set up the relation for the relevant quantity as
\begin{equation}
    \label{derivative_lambdas_relation_s12}
    \begin{aligned}
    &\frac{d^2\langle\langle B_{\text{out}}\rangle\rangle}{d\lambda_1d\lambda_2}\bigg|_{\lambda_1=\lambda_2=0}=\frac{g_2^2g_1\sqrt{\kappa_2}e^{-ct} }{2s_1s_2}\\
    &\quad \times \int_{-\infty}^\infty d\omega \frac{S_{12}(\omega)}{\omega (-is_1-\omega)(-is_2-\omega)},
    \end{aligned}
\end{equation}
where $s_k=i(\Delta_k-\Delta_c)+\frac{\gamma_{dk}}{2}-\frac{\kappa}{2}$.
With the new variable
\begin{equation}
    \label{definition_m12}
    M_{12}(\omega)= \frac{S_{12}(\omega)}{\omega(-is_2-\omega)},
\end{equation}
 we are finally able to write our desired quantity as a convolution with the kernel $K(\Delta_1)$ according to
\begin{equation}
    \label{convolution_gefunden}
    \frac{d^2 \langle \langle B_{\text{out}}\rangle\rangle}{d\lambda_1 d \lambda_2}\bigg|_{\lambda_1=\lambda_2=0}=\frac{g_2^2g_1\pi \sqrt{\kappa_2}e^{-ct}}{s_1s_2}\mathcal{C}(\Delta_1),
\end{equation}
where
\begin{equation}
    \label{definition_der_convolution}
    \begin{aligned}
    \mathcal{C}(\Delta_1)&= (M_{12} \star K)(\Delta_1)\\
    &=\frac{1}{2\pi}\int_{-\infty}^\infty d\omega M_{12}(\omega)K(\Delta_1-\omega)
    \end{aligned}
\end{equation}
is the convolution of $M_{12}(\omega)$ with the kernel
\begin{equation}
    \label{definition_of_the_kernel}
    K( \Delta_1)=\frac{1}{\Delta_1-\Delta_c+\frac{i}{2}(\kappa-\gamma_{d1})}.
\end{equation}
The Fourier transformation of this kernel differs depending on the sign of $\gamma_{d1}-\kappa$. Therefore, we distinguish two cases for the Fourier transformation of the kernel. The respective FT of the kernel yields 
\begin{equation}
    \label{ft_of_kernel_ka_kb}
    \begin{aligned}
         \tilde{K}_{\gamma_{d1>\kappa}}(\tau)&=\frac{1}{2\pi}\int_{-\infty}^\infty d\Delta_1 K(\Delta_1)e^{-i\Delta_1\tau}\bigg|_{\gamma_{d1
         }>\kappa}\\
         &=\bigg\{\begin{array}{cc} ie^{-\big(i\Delta_c+\frac{\kappa-\gamma_{d1}}{2}\big)\tau}, &\text{if} \quad  \tau  \leq 0\\
    0, &\text{if}\quad \tau>0,\end{array} \\
         \tilde{K}_{\gamma_{d1<\kappa}}(\tau)&=\frac{1}{2\pi}\int_{-\infty}^\infty d\Delta_1 K(\Delta_1)e^{-i\Delta_1\tau}\bigg|_{\gamma_{d1}<\kappa}\\
         &=\bigg\{\begin{array}{cc} -ie^{-\big(i\Delta_c+\frac{\kappa-\gamma_{d1}}{2}\big)\tau}, &\text{if} \quad  \tau  \geq 0\\
    0, &\text{if}\quad \tau<0.\end{array} 
    \end{aligned}
\end{equation}
Now we can apply the convolution theorem, which yields 
\begin{equation}
    \label{conv_theo_applied}
    \begin{aligned}
    \tilde{\mathcal{C}}_{\gamma_{d1}>\kappa}(\tau)&=\tilde{M}_{12}(\tau)\tilde{K}_{\gamma_{d1}>\kappa}(\tau), \\  \tilde{\mathcal{C}}_{\gamma_{d1}<\kappa}(\tau)&=\tilde{M}_{12}(\tau)\tilde{K}_{\gamma_{d1}<\kappa}(\tau).
    \end{aligned}
\end{equation}
for the Fourier transformed quantities with the conjugate variable $\tau$ in the two cases. The case separation is necessary, because otherwise the rearrangement of the relation from the convolution theorem for the quantity $\tilde{M}_{12}(\omega)$ is not possible, since this would lead to a division by zero. In order to apply the Fourier back-transformation, it is necessary to consider the case separation, which results in the quantity $M_{12}(\omega)$ with
\begin{equation}
    \label{fourier_back_m12}
    \begin{aligned}
     &M_{12}(\omega)=\int_{-\infty}^\infty d\tau \tilde{M}_{12}(\tau)e^{i\omega \tau} \\
     &=\int_{-\infty}^0 d\tau  \frac{\tilde{C}_{\gamma_{d1}>\kappa}(\tau)}{\tilde{K}_{\gamma_{d1}>\kappa}(\tau)}e^{i\omega \tau}+\int_{0}^\infty d\tau \frac{\tilde{C}_{\gamma_{d1}<\kappa}(\tau)}{\tilde{K}_{\gamma_{d1}<\kappa}(\tau)}e^{i\omega \tau}.
     \end{aligned}
\end{equation}
The next step is to implement the Fourier transformation of the kernels, 
\begin{equation}
    \label{reconstruction_of_m12}
    \begin{aligned}
    M_{12}(\omega)&=-i\int_{-\infty}^0 d\tau  \tilde{\mathcal{C}}_{\gamma_{d1}>\kappa}(\tau)e^{(i(\Delta_c+\omega)+\frac{\kappa-\gamma_{d1}}{2})\tau}\\
    &\quad +i\int_{0}^\infty d\tau  \tilde{\mathcal{C}}_{\gamma_{d1}<\kappa}(\tau)e^{(i(\Delta_c+\omega)+\frac{\kappa-\gamma_{d1}}{2})\tau},
    \end{aligned}
\end{equation}
which yields the noise correlation spectral density together with Eq.~\eqref{definition_m12}.

Another option for calculating the remaining part of $\tilde{M}_{12}(\tau)$ for the half plane, where it cannot be calculated from the convolution theorem because the kernel is zero, consists in applying the residue theorem. In the case where $\gamma_{d1} <\kappa$ (and $\gamma_{d2}<\kappa$ also for symmetry reasons), for example, we can calculate $\tilde{M}_{12}(\tau)$ for $\tau <0$ as
\begin{equation}
    \label{residue_rechnung_a}
    \begin{aligned}
    &\tilde{M}_{12}(\tau)=\frac{1}{2\pi}\int_{-\infty}^\infty d\omega  \frac{S_{12}(\omega)}{\omega(-is_2-\omega)}e^{-i\omega \tau } \\
    &=\frac{1}{2\pi}2\pi i \cdot \frac{\text{Res}_{\omega=0}(M_{12}(\omega)e^{-i\omega \tau})}{2}\\
    &=-\frac{S_{12}(0)}{2s_2}, \ \text{if} \quad \tau <0.
    \end{aligned}
\end{equation}
Using this for the Fourier back-transformation of $M_{12}(\omega)$ yields 
\begin{equation}
    \begin{aligned}
    \label{residuee1}
    &M_{12}(\omega)=\int_0^\infty d\tau \frac{\tilde{C}_{\gamma_{d1}<\kappa}(\tau)}{\tilde{K}_{\gamma_{d1}<\kappa}(\tau)}e^{i\omega \tau}-\frac{\pi S_{12}(0)\delta(\omega)}{2s_2}\\
    &=\int_0^\infty d\tau \frac{\tilde{C}_{\gamma_{d1}<\kappa}(\tau)}{\tilde{K}_{\gamma_{d1}<\kappa}(\tau)}e^{i\omega \tau} \quad \text{if} \quad \omega \ne 0.
    \end{aligned}
\end{equation}
Continuing the calculations by using the second line of this equation is possible, since $M_{12}(\omega)$ is not defined for $\omega=0$ and since this limiting case is not relevant for experiments. In the case, where $\gamma_{d1}> \kappa$ and $\gamma_{d2}>\kappa$, the residue theorem can be applied in the situation where $\tau>0$. This yields a similar result, 
\begin{equation}
    \begin{aligned}
    \label{residuee2}
    &M_{12}(\omega)=\int_{-\infty}^0 d\tau  \frac{\tilde{C}_{\gamma_{d1}>\kappa}(\tau)}{\tilde{K}_{\gamma_{d1}>\kappa}(\tau)}e^{i\omega \tau}+\frac{\pi S_{12}(0)\delta(\omega)}{2s_2}\\
    &=\int_{-\infty}^0 d\tau \frac{\tilde{C}_{\gamma_{d1}>\kappa}(\tau)}{\tilde{K}_{\gamma_{d1}>\kappa}(\tau)}e^{i\omega \tau}, \quad \text{if} \quad \omega \ne 0.
    \end{aligned}
\end{equation}
Comparing these results with Eq.~\eqref{fourier_back_m12} reveals that in the case where $\gamma_{d1}>\kappa$ the contribution of the integral over positive $\tau$ vanishes and in situation, where $\gamma_{d1}<\kappa$, the contribution over negative $\tau$ vanishes. Therefore, and given that $M_{12}(\omega)$ is independent of the sign of $\gamma_{d1}-\kappa$, both contributions are equal to each other in their respective experimental setups, and it is sufficient to consider 
\begin{equation}
    \begin{aligned}
    &M_{12}(\omega)=\int_{0}^\infty d\tau  \frac{\tilde{C}_{\gamma_{d1}<\kappa}(\tau)}{\tilde{K}_{\gamma_{d1}<\kappa}(\tau)}e^{i\omega \tau}\\
    &=i\int_{0}^\infty d\tau \tilde{\mathcal{C}}_{\gamma_{d1}<\kappa}(\tau)e^{(i(\Delta_c+\omega)+\frac{\kappa-\gamma_1}{2})\tau},
    \end{aligned}
\end{equation}
in the situation where $\gamma_{d1}<\kappa$. It should be noted that if this method yields a trivial result for the reconstructed noise spectral density, then the underlying noise spectrum is likely to be dominated by a white noise source. This is due to the fact that the cavity emission has no contribution of the correlations for the white noise case up to third order in the coupling constant $g_k$, as we have shown in Sec.~\ref{sec:white}. 

By neglecting the time-dependent part of the frequency integral, which should in principle also appear in Eq.~\eqref{derivative_lambdas_relation_s12}, an error is introduced. Now we want to estimate this error. The time-dependent parts of this integral, which we neglected, consist of a part that decays exponentially with time $\sim e^{-st}$ and a part that oscillates in time $\sim e^{i\omega t}$. In the limit of $\kappa\ll \gamma_{d1}$ parts $\sim e^{-st}$ decay much faster than the parts with the overall exponential decaying part of $\sim e^{-ct}$, which means that in the limit of high measurement times, these time dependent parts of the integral can be neglected. However, the oscillation terms can not be neglected for longer measurement times. For them, it is necessary to average them out by calculating averages over at least one oscillation period, which enables the possibility of averaging this part out.

\section{Conclusions} 
\label{sec:conclusions}
We demonstrated that an indirect characterization of the noise correlations affecting two qubits is possible from the emission of the cavity to which the qubits are coupled. We investigated the cases of white noise,  quasi-static noise, and Ornstein-Uhlenbeck noise. In the case of white noise, we found extremely small contributions to the cavity emission, which scale with the fifth order in the qubit-cavity coupling. Therefore, these contributions are negligible. Quasi-static correlated noise leads to a change in the cavity emission, which scales approximately linearly in the noise correlation spectral density. In the case of Ornstein-Uhlenbeck noise, the contribution to the noise correlation term shows an exponential decrease with an increasing correlation decay rate (in the time domain), as the noise approaches a white spectrum. For small correlation decay rates, the case of quasi-static noise is recovered, since the noise correlations are scaling approximately linear with the correlation decay rate.

We also provided a prescription for the extraction of the noise correlation decay rate from the cavity emission in the case of OU noise. The comparison between the quasi-static noise case and the Ornstein-Uhlenbeck noise case yields similar results regarding the magnitude of the noise correlation contribution, but for OU noise, the correlation term decays faster. Even in the absence of knowledge about the dominant noise source in a qubit environment, it is still possible to extract the noise correlation spectral density from the cavity emission. For this procedure, it is required to determine the changing rate of the cavity emission with respect to the sensitivity of the two qubits to the noise of the two qubits. This has to be measured as a function of the qubit-probe detuning. Then, the measured quantity can be expressed as a convolution integral. Finally, by applying the convolution theorem, the noise correlation spectral density can be recovered.

The findings of this study can be used to characterize the noisy environment of quantum systems containing multiple qubits.  It is evident that a more sophisticated understanding of the noise can facilitate enhanced performance in the domains of qubit gates and for quantum error correction. It is possible to examine the effects of noise correlations on other parts of the system as well, such as the cavity-qubit coupling.  It would be interesting to explore the noise correlation effects in the regime of strong qubit-cavity coupling. Further interesting aspects of future research are the effects of noise correlations in the context of finite-temperature effects or to study the buildup, as well as degradation, of entanglement between the qubits affected by correlated noise. 

\vspace{1cm}

\section*{Acknowledgments}
\label{sec:Acknowledgments}

This research was supported by the Swiss National Science Foundation (SNSF) through NCCR SPIN (grant no.~225153).
%
\onecolumngrid

\appendix
%
\section{Simplifying the Hamiltonian }
\label{sec:apphamil}
The Hamiltonian for this system can be set up as 
\begin{subequations}
    \label{dicker_fetter_hamiltonian_app}
    \begin{align}
        \mathcal{H}&=\hat{H}_{\text{sys}}(t)+\hat{H}_{\text{bath}}+\hat{H}_{int}, \tag{A1}\\
        \hat{H}_{\text{sys}}(t)&=\omega_c\hat{a}^\dagger\hat{a}+\sum_{k=1,2}\left[\frac{\Delta \omega_{k}+\delta \omega_{k}(t)}{2}\hat{\sigma}_{zk} +\left(g_k\left(\hat{\sigma}_{+k}+\hat{\sigma}_{-k}\right)+g_{lk}\hat{\sigma}_{zk}\right)\left(\hat{a}+\hat{a}^\dagger\right)\right],\\
        \hat{H}_{\text{bath}}&=\sum_{j=1,2}\int_{-\infty}^\infty d\omega  \omega \tilde{\hat{b}}_j^\dagger(\omega)\tilde{\hat{b}}_j(\omega),\\
        \hat{H}_{\text{int}}&=\sum_{j=1,2} \int_{-\infty}^\infty d\omega i\tilde{\kappa}_j(\omega)\left( \tilde{\hat{b}}_j(\omega)\hat{a}^\dagger-\tilde{\hat{b}}_j^\dagger(\omega)\hat{a}\right),
    \end{align}
\end{subequations}
where $\omega_c$ denotes the cavity mode frequency, $\hat{a}^\dagger$ and $\hat{a}$ are the creation and annihilation operators for photons in the cavity. The index $k\in\{1,2\}$ refers to two qubits, while the index $j \in \{1,2\} $ labels the two ports of the cavity. The energy level spacings of the respective qubits are described by $ \Delta\omega_{k}$. The energy levels are affected by longitudinal time-dependent noise $\delta \omega_{k}(t)$, which is described in detail in Sec.~\ref{sec:model_noise}. The terms including Pauli matrices $\hat{\sigma}_{zk}$ and $\hat{\sigma}_{\pm k}=\hat{\sigma}_{xk}\pm i \hat{\sigma}_{yk}$ describe the dynamics of the qubits. The coupling between the qubits and the cavity is composed of a transverse part with a coupling constant $g_k$ and a longitudinal part with a coupling constant $g_{lk}$. The operators $\tilde{\hat{b}}_j^\dagger(\omega)$ and $\tilde{\hat{b}}_j(\omega)$ are frequency-dependent and describe the creation and annihilation of photons outside the cavity ports. Finally, $\tilde{\kappa}_j(\omega)$ is the coupling strength between the cavity ports and the thermal bath, i.e., the surroundings of the cavity. Note that we set $\hbar=1$ throughout this work. In order to simplify the model, a transformation into a rotating frame is performed according to
\begin{equation}
    \label{unitary_transform_app}
    \mathcal{H}(t)\rightarrow\hat{U}\mathcal{H}(t)\hat{U}^\dagger +i\left(\partial_t\hat{U}\right)\hat{U}^\dagger = \mathcal{H}_r,
\end{equation}
 with the unitary 
\begin{equation}
    \label{unitary_app}
    \hat{U}=\text{exp}\left[ i\omega_p t\left(\frac{\hat{\sigma}_{z1}+\hat{\sigma}_{z2}}{2}+\hat{a}^\dagger\hat{a}\right)\right].
\end{equation}
The Hamiltonian in its rotating frame ca be stated as,
\begin{subequations}
\allowdisplaybreaks
    \label{hamil_rot_frame_app}
    \begin{align}
        &\mathcal{H}_r=\hat{H}_\text{sys,r}(t)+\hat{H}_{\text{bath}}+\hat{H}_{\text{int,r}}, \tag{A4}\\
        &\hat{H}_{\text{sys,r}}(t)=\Delta_c \hat{a}^\dagger\hat{a}+\sum_{k=1,2}\bigg[\frac{\Delta_k+\delta\omega_{k}(t)}{2}\hat{\sigma}_{zk} +g_{lk}\hat{\sigma}_{zk}\left[e^{-i\omega_pt}\hat{a}+e^{i\omega_pt}\hat{a}^\dagger \right] \notag  \\
        &\quad +g_k\bigg( \hat{\sigma}_{+k}\hat{a} +\hat{\sigma}_{+k}\hat{a}^\dagger e^{2i\omega_pt}+\hat{\sigma}_{-k}\hat{a}e^{-2i\omega_p t }+\hat\sigma_{-k}\hat{a}^\dagger\bigg)\bigg],\\ 
        &\hat{H}_{\text{bath}}=\sum_{j=1,2}\int_{-\infty}^\infty d\omega \omega \hat{b}_j^\dagger(\omega)\hat{b}_j(\omega),\\
        &\hat{H}_{\text{int,r}}=\sum_{j=1,2}\int_{-\infty}^\infty d\omega i\tilde{\kappa}_j(\omega)\left( \hat{b}_j(\omega)\hat{a}^\dagger-\hat{b}_j^\dagger(\omega)\hat{a}\right),
    \end{align}
\end{subequations}
where the new operators are defined as $\hat{b}_j(\omega)=\tilde{\hat{b}}_j(\omega)e^{i\omega_p t}$ and $\hat{b}_j^\dagger(\omega)=\tilde{\hat{b}}_j^\dagger(\omega)e^{-i\omega_pt}$. The qubit-probe detunings for the respective qubits $k \in\{1,2\}$ denoted by $\Delta_k=\Delta\omega_{k}-\omega_p$ and the cavity-probe detuning designated as $\Delta_c=\omega_c-\omega_p$ are also introduced. The Hamiltonian part describing the bath $\hat{H}_{\text{bath}}$ remains unaffected by the transformation into the rotating frame. 

In the following, we perform a rotating wave approximation (RWA), which consists of neglecting the fast rotating terms $\sim e^{\pm2i\omega_p t }$. This approximation only affects the system part of the Hamiltonian, $\hat{H}_{\text{sys,r}}$. After applying the RWA, the final form of the Hamiltonian is
\begin{subequations}
    \label{hamiltonian_mit_rwa_app}
    \begin{align}
         \mathcal{H}_R&=\hat{H}_\text{sys,R}(t)+\hat{H}_{\text{bath}}+\hat{H}_{\text{int,r}}, \tag{A5}\\
        \hat{H}_{\text{sys,R}}(t)&=\Delta_c \hat{a}^\dagger\hat{a}+\sum_{k=1,2}\left[\frac{\Delta_k+\delta\omega_{k}(t)}{2}\hat{\sigma}_{zk} +g_k\left( \hat{\sigma}_{+k}\hat{a}+\hat\sigma_{-k}\hat{a}^\dagger\right)\right].\label{5a}
    \end{align}    
\end{subequations}
In the main text, we drop the indices $R$ and $r$ for simplification.


\section{QLE Simplification}
\label{sec:appqle}
We derive the equation of motion for the relevant system operators by applying Eq.~\eqref{quantum_master_equation} to the operators $\hat{\sigma}_{-k}$ and $\hat{\sigma}_{zk}$. This set of differential equations which is obtained thereby is known as the Quantum Langevin equations (QLE),
\begin{subequations}
\label{QLE_original_app}
\allowdisplaybreaks
    \begin{align}
        \frac{d\langle\hat{\sigma}_{-k}\rangle}{dt}&=\bigg(-i(\Delta_k+\delta\omega_{k}(t))-\frac{\gamma_{d k}}{2}\bigg)\langle\hat{\sigma}_{-k}\rangle+ig_k\langle\hat{\sigma}_{zk}\hat{a}\rangle, \label{16aapp}\\
        \frac{d\langle \hat{a}\rangle}{dt}&=-i\Delta_c\langle\hat{a}\rangle-i\sum_{k=1,2}g_k\langle\hat{\sigma}_{-k}\rangle +\sum_{j=1,2}\int_{-\infty}^\infty d\omega \tilde{\kappa}_j(\omega)\langle\hat{b}_j(\omega)\rangle ,\label{16bapp}\\
        \frac{d\langle \hat{b}_j(\omega)\rangle}{dt}&=-\tilde{\kappa}_j(\omega)\langle\hat{a}\rangle-i\omega \langle \hat{b}_j(\omega)\rangle , \label{16capp}\\
        \frac{d\langle \hat{\sigma}_{zk}\rangle }{dt}&=-\gamma_k-(1+2n_k)\gamma_k\langle \hat{\sigma}_{zk}\rangle  +2ig_k\left( \langle \hat{a}^\dagger\hat{\sigma}_{-k}\rangle-\langle \hat{a}\hat{\sigma}_{+k}\rangle\right),\label{16dapp}
    \end{align}
\end{subequations}
where we defined the total noise-independent decoherence rate as $\gamma_{dk}=(1+2n_k)\gamma_k+2\gamma_{\phi k}$. Next, we simplify the QLE. First, the frequency dependence of the coupling between the cavity and its surroundings is neglected, $\tilde{\kappa}(\omega)=\tilde{\kappa}$. Then, Eq.~\eqref{16capp} is solved with
\begin{equation}
    \label{solution_of_16c_app}
    \langle\hat{b}_j(\omega)\rangle=\left[\langle \hat{b}_{j,0}(\omega)\rangle -\int_0^t dt'\tilde{\kappa}_j\langle\hat{a}(t')\rangle e^{i\omega t'}\right]e^{-i\omega t},
\end{equation}
where $\langle \hat{b}_j(\omega)\rangle(t=0)=\langle\hat{b}_{j,0}(\omega)\rangle$. This solution can be substituted into Eq.~\eqref{16bapp}, which gives 
\begin{equation}
    \label{zwischenschritt_app}
    \begin{aligned}
        &\frac{d\langle\hat{a}\rangle}{dt}=-i\Delta_c\langle\hat{a}\rangle-i\sum_{k=1,2}g_k\langle\hat{\sigma}_{-k}\rangle +\sum_{j=1,2}\int_{-\infty}^\infty d\omega  \tilde{\kappa}_j\bigg(\langle\hat{b}_{0,j}(\omega)\rangle -\int_0^t dt'\tilde{\kappa}_j\langle\hat{a}(t')\rangle e^{i\omega t'}\bigg)e^{-i\omega t}.
    \end{aligned}
\end{equation}
After evaluating the frequency integral with $\int_{-\infty}^\infty e^{-i\omega(t-t')}d\omega=2\pi \delta(t-t')$ and the time integral by applying $\int_0^tc(t')\delta(t-t')=\frac{1}{2}c(t)$, the differential equation for $\langle\hat{a}\rangle$ can be obtained as
\begin{equation}
    \label{qleq_for_a_simplified_app}
    \begin{aligned}
    \frac{d \langle \hat{a}\rangle}{dt}&=\left(-i\Delta_c-\frac{\kappa}{2}\right)\langle \hat{a}\rangle-i\sum_{k=1,2}g_k\langle \hat{\sigma}_{-k}\rangle +\sum_{j=1,2}\sqrt{\kappa_j}\langle\hat{b}_{in,j}(t)\rangle,
    \end{aligned}
\end{equation}
where the input field at port $j$ is defined as 
\begin{equation}
    \label{definition_bin_app}
    \langle \hat{b}_{in,j}(t)\rangle=\frac{1}{\sqrt{2\pi}}\int_{-\infty }^\infty d\omega e^{-i\omega t}\langle \hat{b}_{j,0}(\omega)\rangle .
\end{equation}
The couplings between the cavity and the environment are redefined as $\kappa=\kappa_1+\kappa_2$ and $\kappa_j=2\pi \tilde{\kappa}_j^2$ \cite{inoutgardiner}.

To further simplify the system of differential equations, some assumptions are introduced. First, in order to describe the noise correlation effects from the cavity emission arising from the relaxation of the qubits into their ground states, after preparation in a superposition state, the case without input fields at both ports is considered  $\langle \hat{b}_{in,j}(t)\rangle=0$. In addition, the initial state is assumed to be separable, i.e., without any entanglement between the cavity and the qubits. This leads to $\langle \hat{X}\hat{Y}\rangle=\langle\hat{X}\rangle\langle \hat{Y}\rangle$. Next, for the environment, the low-temperature limit is considered $T\ll\Delta \omega_{k}$, where the average number of particles in the thermal bath is $n_k=0$, thus neglecting any stimulated emission and absorption processes.   With these approximations, the final set of the remaining QLE, Eq.~\eqref{final_set_of_qleqs}, is obtained.

\section{Cavity emission for quasi-static noise}
\label{sec:app_qs_formulas}
In the fully symmetric quasi-static noise case, the autocorrelation contribution to the noise-correction part of the cavity emission can be stated as follows,
\begin{equation}
    \label{output_field_for_qs_noise_in_tse_APP}
    \begin{aligned}
        &\langle\langle B_\text{out}^{qk}(t)\rangle\rangle_{\text{qs}}=-\frac{i\langle\hat{\sigma}_{-,0}\rangle g S_q \sqrt{\kappa_2}\lambda^2e^{-ct}}{2s^3}\bigg( 2 +e^{-st}(-2-st(2+st)) +g^2\bigg(\frac{1}{s^2}\bigg( 24-6st -e^{-st}(24 +st(18\\
        &\quad +st(6+st)))\bigg)
         -\frac{(\langle\hat{\sigma}_{z,0}\rangle+1)}{\gamma s_{-}^3}\bigg(2e^{-\gamma t}(3s^2-3s\gamma+\gamma^2) -e^{-st}(s^2(6+st(4+st))-2s(3+st(3+st))\gamma\\
         &\quad +(2+st(2+st))\gamma^2)\bigg) +\frac{(\langle\hat{\sigma}_{z,0}\rangle+1)}{\gamma s_{+}^3}\bigg(2(1 -e^{-s_{+}t})(3s^2+3s\gamma+\gamma^2)-e^{-s_{+}t}sts_{+}(2\gamma +s(4+ts_{+}))\bigg)\bigg)\bigg).
    \end{aligned}
\end{equation}
When both qubits couple to the same cavity, an additional contribution to the autocorrelation noise correction term appears, which is not present in the case, where an individual qubit coupled to a cavity. This contribution is given by
\begin{equation}
    \label{b_qk+_add_term_qs_APP}
    \begin{aligned}
    &\langle\langle B_{\text{out}}^{\text{qk+}}(t)\rangle\rangle_{\text{qs}}=-\frac{i\langle\hat{\sigma}_{-,0}\rangle g^3S_q \lambda^2 \sqrt{\kappa_2}e^{-ct} }{s^3}\bigg(\frac{1}{s^2}\bigg(16-4st-2e^{-st}\bigg(8+st\bigg(6+st\bigg(2+\frac{st}{3}\bigg)\bigg)\bigg)\bigg)\\
    &+\frac{\langle\hat{\sigma}_{z,0}\rangle+1}{s_{-}^3\gamma^3}\bigg(-4ss_{-}^3-2e^{-\gamma t}\gamma^2(2s^2-2s\gamma+\gamma^2)+e^{-st}(4s^4-2s^3(6+st)\gamma+s^2(4+st)^2\gamma^2\\
    &-2s(2+st)^2\gamma^3+(2+st(2+st))\gamma^4)\bigg)+\frac{\langle\hat{\sigma}_{z,0}\rangle+1}{s_{+}^3\gamma^3}\bigg(2(2s^4+6s^3\gamma+8s^2\gamma^2+4s\gamma^3+\gamma^4)\\
    &+e^{-s_{+}t}(-4s^4-2s^3(6+st)\gamma-s^2(4+st)^2\gamma^2-2s(2+st)^2\gamma^3-(2+st(2+st))\gamma^4\bigg)\bigg)\bigg).
    \end{aligned}
\end{equation}
Finally, the cross-correlation term, from which the noise spectral density $S_{12}(\omega)$ can be extracted, is given by
\begin{equation}
    \label{qs_correlation_terms_qs_APP}
    \begin{aligned}
        &\langle\langle B_{\text{out} }^{q1,q2}(t)\rangle\rangle_{\text{qs}}=\frac{2i\langle\hat{\sigma}_{-,0}\rangle g^3S_{12}\lambda^2\sqrt{\kappa_2}e^{-ct}}{s^2}\bigg( \frac{-4+s t}{s^3}-\frac{\langle\hat{\sigma}_{z,0}\rangle+1}{s_{+}^2\gamma}+\frac{\langle\hat{\sigma}_{z,0}\rangle+1}{\gamma^3}\bigg(\frac{\gamma^2e^{-\gamma t}}{s_{-}^2}\\
        &-\frac{e^{-s_{+}t}\big(2s^2+4s\gamma+ss_{+}t\gamma+\gamma^2\big)}{s_{+}^2}\bigg)+e^{-st}\bigg(\frac{4+st(3+st(1+\frac{st}{6}))}{s^3}+\frac{\langle\hat{\sigma}_{z,0}\rangle+1}{s_{-}^2\gamma^3}\bigg(\gamma^2\\
        &+s^2(2-\gamma t)+s\gamma(-4+\gamma t)\bigg)\bigg)\bigg).
    \end{aligned}
\end{equation}

\section{Cavity emission for Ornstein-Uhlenbeck noise}
\label{sec:app_OU_formulas}
The autocorrelation contribution to the noise correction part of the cavity emission can be stated as 
\begin{subequations}
    \allowdisplaybreaks
    \label{auto_corr_ou_linearer_term_APP}
    \begin{align}
    &\langle\langle B_{\text{out}}^{\text{qk}}(t)\rangle\rangle_{\text{OU}}=\langle\langle B_{\text{out},g^1}^{\text{qk}}(t) +  B_{\text{out},g^3}^{\text{qk}}(t)\rangle\rangle_{\text{OU}}, \tag{D1}\\
    &\langle\langle B_{\text{out},g^1}^{\text{qk}}(t)\rangle\rangle_{\text{OU}}=\frac{ig\langle\hat{\sigma}_{-,0}\rangle\sqrt{\kappa_2}\lambda^2}{\Gamma^2}\bigg( \frac{e^{-(q+\Gamma)t}-e^{-ct}}{s+\Gamma} +\frac{e^{-ct}(s-\Gamma)+e^{-qt}(\Gamma-s+s\Gamma t)}{s^2}\bigg),\\
    &\langle\langle B_{\text{out},g^3}^{\text{qk}}(t)\rangle\rangle_{\text{OU}}=\sqrt{\kappa_2}e^{-ct}\int_0^{t}dt'ig^3e^{-st'}  \int_0^{t'}dt''\big(-1+\left( \langle \hat{\sigma}_{z,0}\rangle+1\right)e^{-\gamma t''}\big) e^{st''}\langle\hat{\sigma}_{-,0}\rangle\notag \\
    &\quad \times \int_0^{t''}dt'''e^{-st'''} \frac{\lambda^2}{\Gamma^2}\bigg[2-\Gamma(t'-t''+t''') -e^{-\Gamma t'''}  +e^{-\Gamma t''}-e^{-\Gamma t'}-e^{\Gamma (t''-t')}-e^{\Gamma (t'''-t'')} +e^{\Gamma (t'''-t')}\bigg].
    \end{align}
\end{subequations}
in the case of OU noise. Finally, the noise correlation contribution to the cavity transmission is given by

\begin{equation}
    \label{noi_corr_for_ou_term_APP}
    \begin{aligned}
    &\langle\langle B_{\text{out}}^{\text{q1,q2}}(t)\rangle\rangle_{\text{OU}}=\frac{2ig^3 \langle\hat{\sigma}_{-,0}\rangle \sqrt{\kappa_2}\lambda^2}{s \Gamma_{12}^2(s-\Gamma_{12})}\bigg[\frac{s \Gamma_{12}(\langle\hat{\sigma}_{z,0}\rangle+1)}{\gamma s_{+}}\bigg(\frac{e^{-(q+\gamma)t}}{\Gamma_{12}-\gamma}-\frac{2se^{-ct}}{s_{-}(s+\Gamma_{12})}\bigg)+\frac{e^{-(q+\Gamma_{12})t}(\Gamma_{12}-s)}{s\Gamma_{12}(s+\Gamma_{12})^2}\bigg(-2s^2\\
    &\quad-4s\Gamma_{12}-\Gamma_{12}^2-s\Gamma_{12}(s+\Gamma_{12})t+\frac{s^2}{s_{-}}(\Gamma_{12}(s+\Gamma_{12})\alpha_{-})\bigg)+ \frac{\Gamma_{12}\big(e^{-ct}(\Gamma_{12}-s)^2-e^{-(c+\Gamma_{12})t}(s+\Gamma_{12})^2\big) }{s(s-\Gamma_{12})(s+\Gamma_{12})^2}   
    \\
    &\quad -\frac{\alpha_{+}\Gamma_{12}\gamma \big(e^{-(c+\gamma+\Gamma_{12})t}-e^{-ct}\big)}{s_{-}(\Gamma_{12}-s_{-})}+e^{-qt} \bigg(\frac{1}{s}+\frac{1}{s-\Gamma_{12}}-\frac{2}{\Gamma_{12}}+t-\frac{\alpha_{+}(\Gamma_{12}-s)}{ \Gamma_{12}-s_{-}}\bigg)\\
    &+\frac{\alpha_{+} (s-\Gamma_{12})}{(s_{+}+\Gamma_{12})(\Gamma_{12}-s_{-})}\bigg(e^{-(q+\gamma+\Gamma_{12})t}(\Gamma_{12}-s_{-})-2e^{-ct}(\Gamma_{12}+\gamma)\bigg)\bigg].
    \end{aligned}
\end{equation}

\twocolumngrid
\bibliography{biblio.bib}

\end{document}